\documentclass{elsart5p}
\pdfoutput=1
 \usepackage{graphics}
 \usepackage{graphicx}
\usepackage{subfigure}
\usepackage{algorithm}
\usepackage{algorithmic}
\usepackage{times}
\usepackage{verbatim}
\usepackage{ifpdf}
\usepackage{amsmath} 
\usepackage{amsxtra}
\usepackage{amssymb}
\usepackage{latexsym}
\usepackage{graphicx} 

\begin{document}

\begin{frontmatter}

\title{Characterizing the Robustness of Complex Networks}

\author{Ali Sydney, Caterina Scoglio, Mina Youssef, \and Phillip Schumm}

\address{Electrical and Computer Engineering Department\\ Kansas State University \\ Manhattan, KS USA\\Email:\{asydney,caterina,mkamel,pbschumm\}@ksu.edu}

\begin{abstract}
With increasingly ambitious initiatives such as GENI and FIND that seek to
design future internets, it becomes imperative to define the
characteristics of robust topologies, and build future networks optimized
for robustness. This paper investigates the characteristics of network
topologies that maintain a high level of throughput in spite
of multiple attacks. To this end, we select network topologies belonging
to the main network models and some real world networks. We consider three
types of attacks: removal of random nodes, high degree nodes, and high
betweenness nodes. We use elasticity as our robustness measure and, through our analysis, illustrate that different
topologies can have different degrees of robustness. In particular, elasticity can fall as low as $0.8\%$ of the upper bound based on the attack employed. This result substantiates the need for
optimized network topology design. Furthermore, we implement a tradeoff function that combines elasticity under the three attack strategies and considers the cost of the network. Our extensive simulations
show that, for a given network density, regular and semi-regular
topologies can have higher degrees of robustness than heterogeneous
topologies, and that link redundancy is a sufficient but not necessary
condition for robustness.  
\end{abstract}

\end{frontmatter}

\section{Introduction}
Why study future network topologies?  For one, we have experienced several moderate sized failures and thus, large failures are inevitable. In particular, the 2006 earthquake in Taiwan disrupted undersea fiber optic communication lines and as a result, banks from South Korea to Australia suffered massive interruptions \cite{earthquake}. Though this represents a direct network failure, failures can also occur indirectly. For example Code Red, a computer virus that incapacitated numerous networks, resulted in a global loss of 2 billion US dollars \cite{codered}. Furthermore, in 2004, Sassar virus disruptions accounted for the halt on maritime operations in the UK, the halt on railway operations in Australia, and interruptions in hospital facilities in Hong Kong \cite{sassar}.  The US General Accounting Office estimated 250,000 annual attacks on Department of Defense networks \cite{dod}.  Objectives range from theft to immobilization of entire networks. Another riveting example stems from a series of cascading failures in 2003 that resulted in a blackout in the Northeastern states \cite{blackout}. A similar phenomenon occurred the very same year in Italy, and left 56 million residents without power for 9 hours \cite{italy}.   Our daily routines would cease to exist should network topologies disintegrate.  Thus, as failures and attacks increase, it is imperative to design future topologies robust against unforeseen catastrophes for future network initiatives. 

Amongst other definitions, a network can be robust if disconnecting components is difficult. However, we define robustness as the ability of a network to maintain its total throughput under node and link removal. The former definition is based on topological characteristics, while the latter also considers flows within the network such as IP packets.  

Approaches for determining the robustness of graphs have evolved from simple graph theoretic concepts that highlight the connectivity of a graph \cite{HIT70} to more recent concepts that consider the spectrum of a graph \cite{as07}.  However, these measures are unable to capture our definition of robustness.  For this reason, we use elasticity as our measure of robustness; it meets the functional requirements of capturing throughput under node and link removal.

The importance of this paper stems from our objective to extract the characteristics of robust networks.  With these results, we seek to produce future robust network topologies. Thus, to realize our first goal, we 1) use the metric elasticity as a measure of robustness of a network, 2) establish the upper bound for elasticity, 3) assess elasticity for diverse network models, 4) present correlations between elasticity and selected network metrics, 5) develop and implement a function that considers the tradeoff between elasticity and network cost, and 6) extract characteristics of networks that make them robust.

The rest of this paper is structured as follows.  Section \ref{background} reviews measures of robustness based on the structure and behavior of the network.  Section \ref{networkmodels} presents the network models from which networks will be selected to assess their elasticity.  In Section \ref{robustness}, we review elasticity, our robustness measure, and provide analytical and numerical approaches to obtain the upper bound.  In Section \ref{experimentalresults}, we assess the elasticity of each network, implement a tradeoff function that considers elasticity under the three removal strategies and discuss the characteristics which make a network robust. Finally, we discuss the benefits and shortcomings of elasticity and highlight our future initiatives to characterize the robustness of complex networks in Section \ref{conclusions}.

\section{Background and Related Work}
\label{background}

The classical approach for determining robustness of networks
entails the use of basic concepts from graph theory. For instance,
the connectivity of a graph is an important, and probably the
earliest, measure of robustness of a network \cite{HIT70}. Node (link)
connectivity, defined as the size of the smallest node (link) cut,
determines in a certain sense the robustness of a graph to the
deletion of nodes (links). However, the node or link connectivity only
partly reflects the ability of graphs to retain certain degrees
of connectedness after deletion. Other improved measures were
introduced and studied, including super connectivity \cite{DFCR81},
conditional connectivity \cite{F83}, restricted connectivity \cite{AHSL88}, fault
diameter \cite{MSB87}, toughness \cite{V73}, scattering number \cite{HA78}, tenacity \cite{MDS95},
expansion parameter \cite{N86}, and isoperimetric number \cite{M89}. In
contrast to node (link) connectivity, these new measures consider
both the cost to damage a network and how badly the network is
damaged. 

Subsequent measures consider the size of the largest connected component as nodes are attacked \cite{rhal00}. Furthermore, percolation models were used to assess the damage incurred by random graphs \cite{AMA08}. From spectral analysis, experimentalists consider the second smallest Laplacian eigenvalue as a measure of how difficult it is to break the network into components \cite{as07}. 

The measures reviewed thus far consider the network structure to assess robustness. However, more recent efforts have incorporated the behavior of the network \cite{dlwc05,ACPRE08}. More precisely, the authors maximized flows in the network while imposing constraints on routers and links. 

Other metrics in networking literature include the average node degree \cite{pdmbxkca05},  betweenness \cite{PDK06}, heterogeneity \cite{JS07}, and characteristic path length \cite{dl08}. In this paper, our results show significant corelations between elasticity and some of these metrics which will be used to characterize the robustness of networks.

\section{Network Models}
\label{networkmodels}
This section reviews the six models from which 18 topologies were selected. They include networks from random models, Watts-Strogatz models, preferential attachment models, near-regular models, trade-off and optimization models, and real-world models. For each topology, some of the more common properties are shown in Table \ref{tab:netchar1}.

\begin{table}[h]
\centering
\caption{Network characteristics where ASP is the average shortest path and Het is heterogeneity}
    \begin{tabular}[c]{|c|c|c|c|c|c|c|}
        \hline
        {\bf Networks }		& 	{\bf $\sharp$ Nodes} 		& {\bf $\sharp$ Links} 		&  {\bf Density} 		& {\bf Diameter} 		& {\bf ASP} &  {\bf Het}\\
        \hline  
				Gi-dense					&	1000	&	4505	&	0.00902	&	7	&	3.391			&	0.331 \\ \hline
				MySpace						&	955	&	10976	&	0.02409	&	4	&	2.013						&	2.027 \\ \hline
				Watts-Strogatz 1	&	1000	&	3000	&	0.00601	&	7	&	4.14						&	0.301	\\ \hline
				PA 2							&	1000	&	2964	&	0.00593	&	6	&	3.534				&	1.109	\\ \hline
				Gi-sparse					&	1000	&	2009	&	0.00402	&	12	&	5.154			&	0.491	\\ \hline
				PA 1							&	1000	&	1981	&	0.00397	&	8	&	4.177		&	1.185	\\ \hline
				Watts-Strogatz 2	&	1000	&	2000	&	0.004	&	9	&	5.294					&	0.37	\\ \hline
				YouTube						&	1089	&	1576	&	0.00266	&	12	&	5.096		&	1.319	\\ \hline
				Flickr						&	967		&	1515	&	0.0032	&	12	&	4.624				&	1.394	\\	\hline
				meshcore					&	1000	&	1275	&	0.00255	&	3	&		2.911			&	3.796	\\ \hline
				near-regular 2		&	992	&	3781	&	0.00769	&	31	&		14.706				&	0.133	\\ \hline
				HOT 2							&	1000			&	1049	&	0.0021	&	12	&	7.144		&	1.892	\\ \hline
				ringcore				&	1000	&	1000	&	0.002	&	14	&	8.196			&	3.122 \\ \hline
				HOT 1							&	939	&	988	&	0.00224	&	10	&		6.812				&				2.032 \\ \hline
				PA-sparse					&	1000	&	1049	&	0.0021	&	14	&	5.793		&	1.892 \\ \hline
				Abilene						&	886	&	896	&	0.00229	&	10	&	6.95		&	2.09 \\ \hline
				near-regular 1	&	992	&	1921	&	0.00391	&	61	&		21			&0.089     \\ \hline    
      
        \hline
    \end{tabular}
    
    \label{tab:netchar1}

\end{table}

\subsection{Random models}
\begin{figure}[h]
\centering  
    \includegraphics[height=2in]{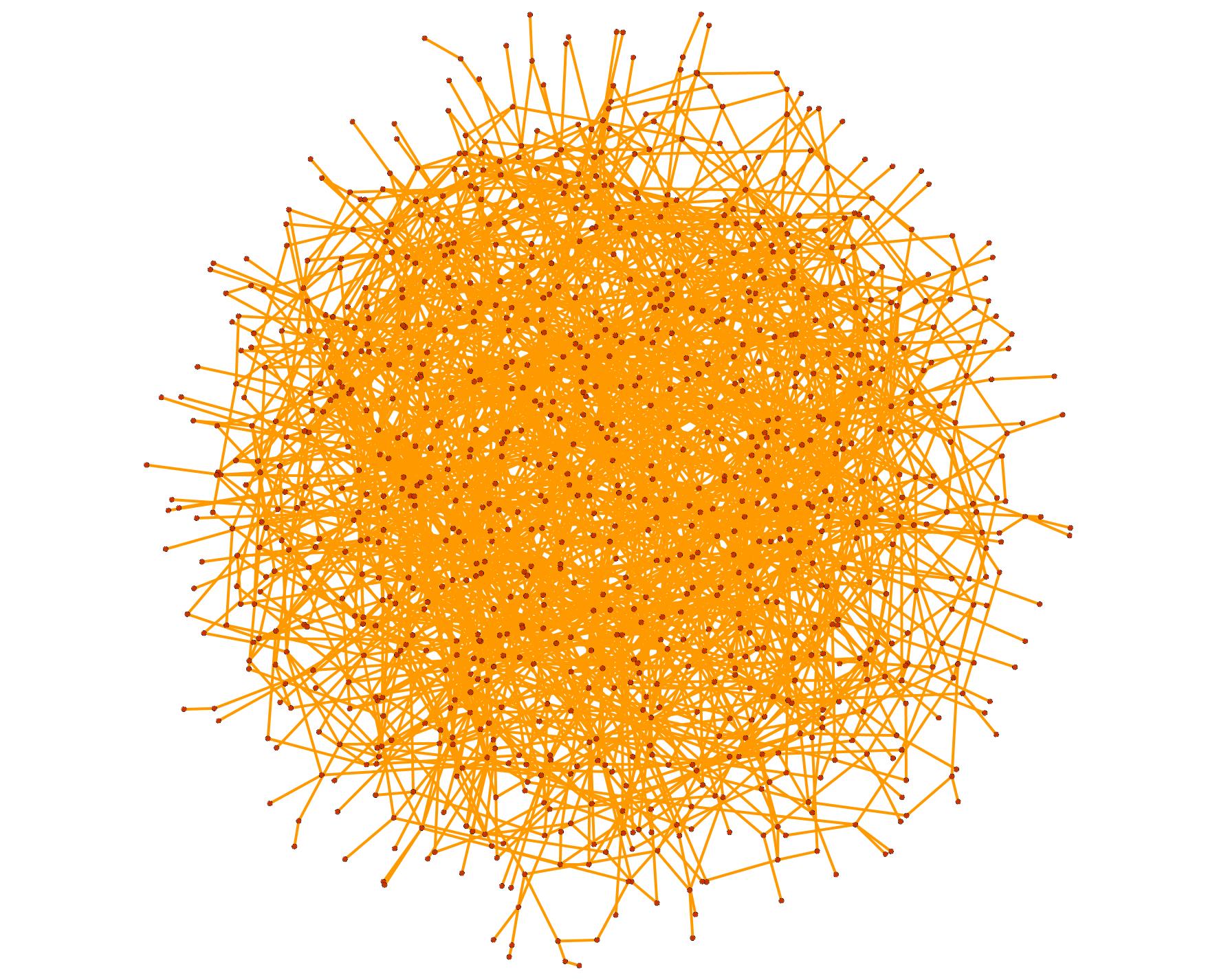}
    \caption[Lattice]{The Gi-sparse network with size $N = 1000$ and average degree $\bar{k} = 4.018$}
    \label{fig:Gi-sparse}      
\end{figure}

A random graph is obtained by random addition of links between $n$ vertices.  Two notable properties are 1) the average node degree determines the connectivity of the graph and 2) the node degree can be approximated using a Poisson distribution.  Erdos-Renyi's (ER) stochastic model is one of the most studied of these models.  In the construction of an ER graph $G(N,E)$, $E$ edges are connected at random to $N$ nodes \cite{AMA08}.   However, this paper considers the Gilbert (Gi) model $G(N,p)$, a modified version of the ER model where edges are connected to vertices with a probability of $p$. For the Gi-dense and Gi-sparse networks used in this paper, $p=0.0091$ and $0.004094$ respectively \cite{NWBTEAM}. Figure \ref{fig:Gi-sparse} shows the Gi-sparse network.

\subsection{Watts-Strogatz Models}
\begin{figure}[h]
\centering  
    \includegraphics[height=2in]{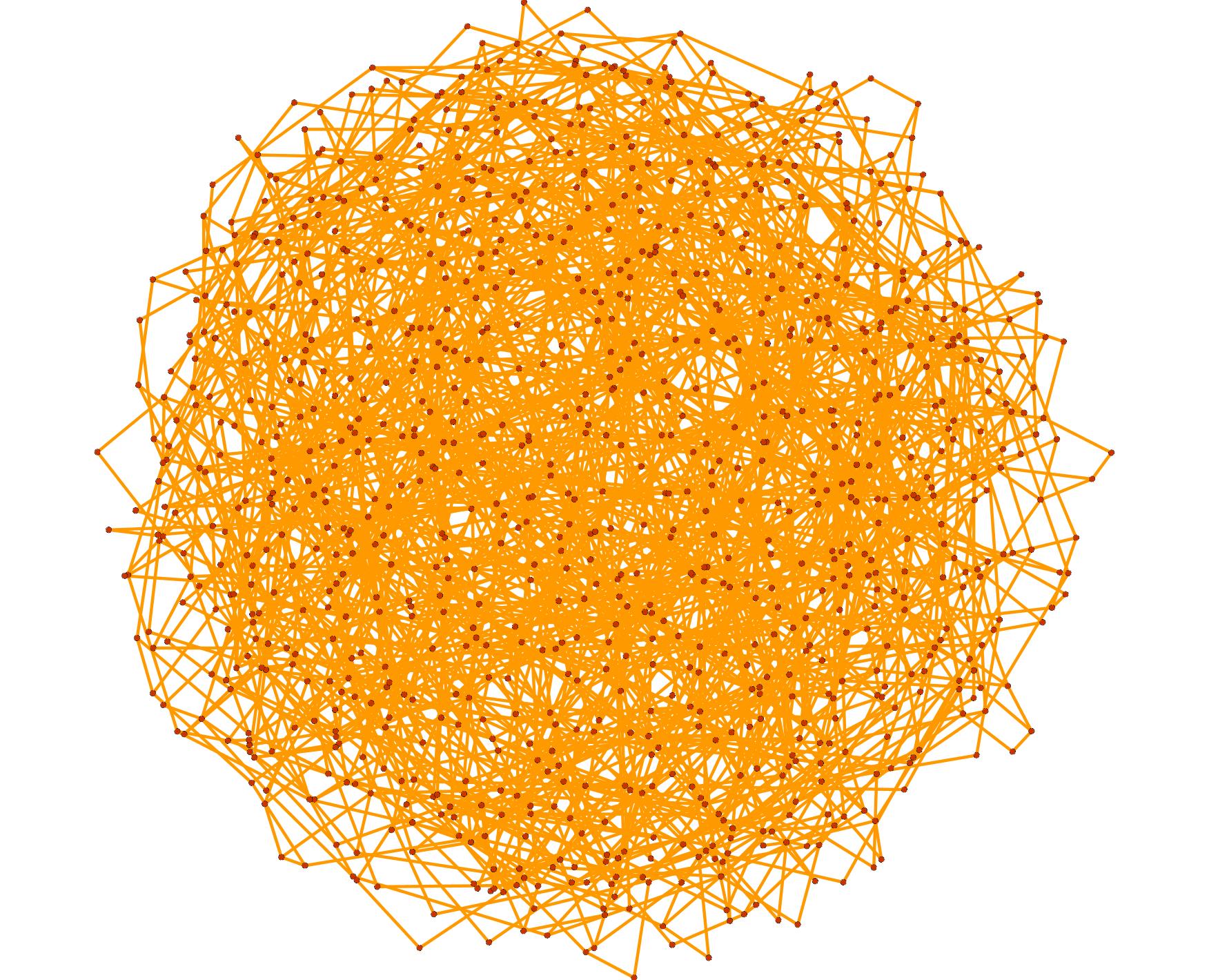}
    \caption[Lattice]{The W-S 1 network with $N = 1000$ and $\bar{k} = 4$ }
    \label{fig:W-S1}      
\end{figure}

The Watts-Strogatz model is constructed by interpolating between a regular ring lattice and a random network \cite{AMA08}. Each node is connected to its $k$ nearest neighbors and random rewiring occurs with a probability of $p$. For intermediate values of $p$, Watts-Strogatz models produces a Small-world network which captures the high clustering properties of regular graphs and the small characteristic path length of random graph models.  For the Watts-Strogatz (W-S) 1 and 2 networks used, the rewiring probability was $0.3$ and $0.5$ \cite{NWBTEAM}. Figure \ref{fig:W-S1} shows the W-S 1 network.

\subsection{Preferential Attachment Models}
\begin{figure}[h]
\centering  
    \includegraphics[height=2in]{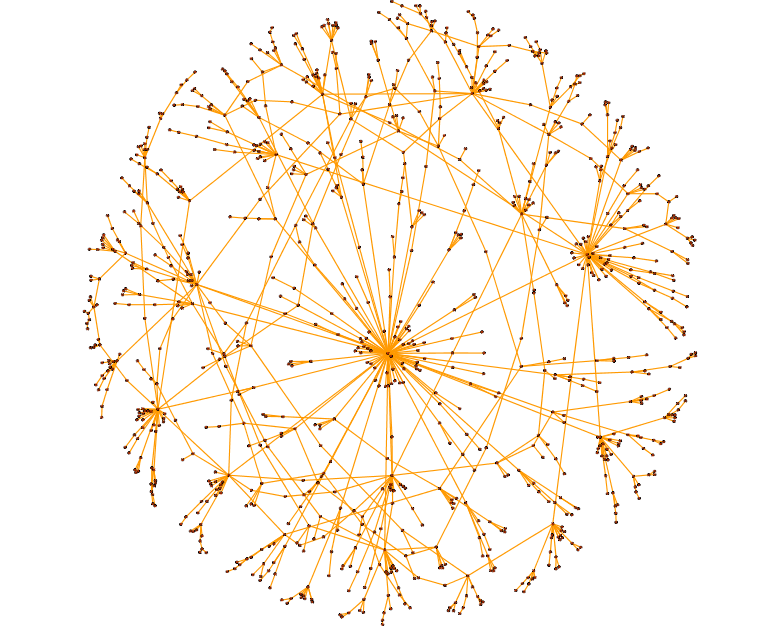}
    \caption[Preferential attachment]{The PA-sparse network with $N = 1000$ and $\bar{k}=2.098$}
    \label{fig:pasparse}      
\end{figure}

From their origin, preferential attachment (PA) models have been considered vulnerable to targeted attacks while robust to random failures and have a heavy tail distribution \cite{ALE03}.  This model constitutes popular nodes called ``hubs" that have a large number of neighbors compared to other nodes with few neighbors. At each time step, nodes with a higher degree have a higher probability of attracting new nodes than nodes with a lower degree.  For this work, the PA 1, PA 2, and PA-sparse networks were constructed using the Barabasi-Albert Scale-free model \cite{NWBTEAM,dlwc05}. Figure \ref{fig:pasparse} shows the PA-sparse network.

\subsection{Near-Regular Models}
\begin{figure}[h]
\centering  
    \includegraphics[height=2in]{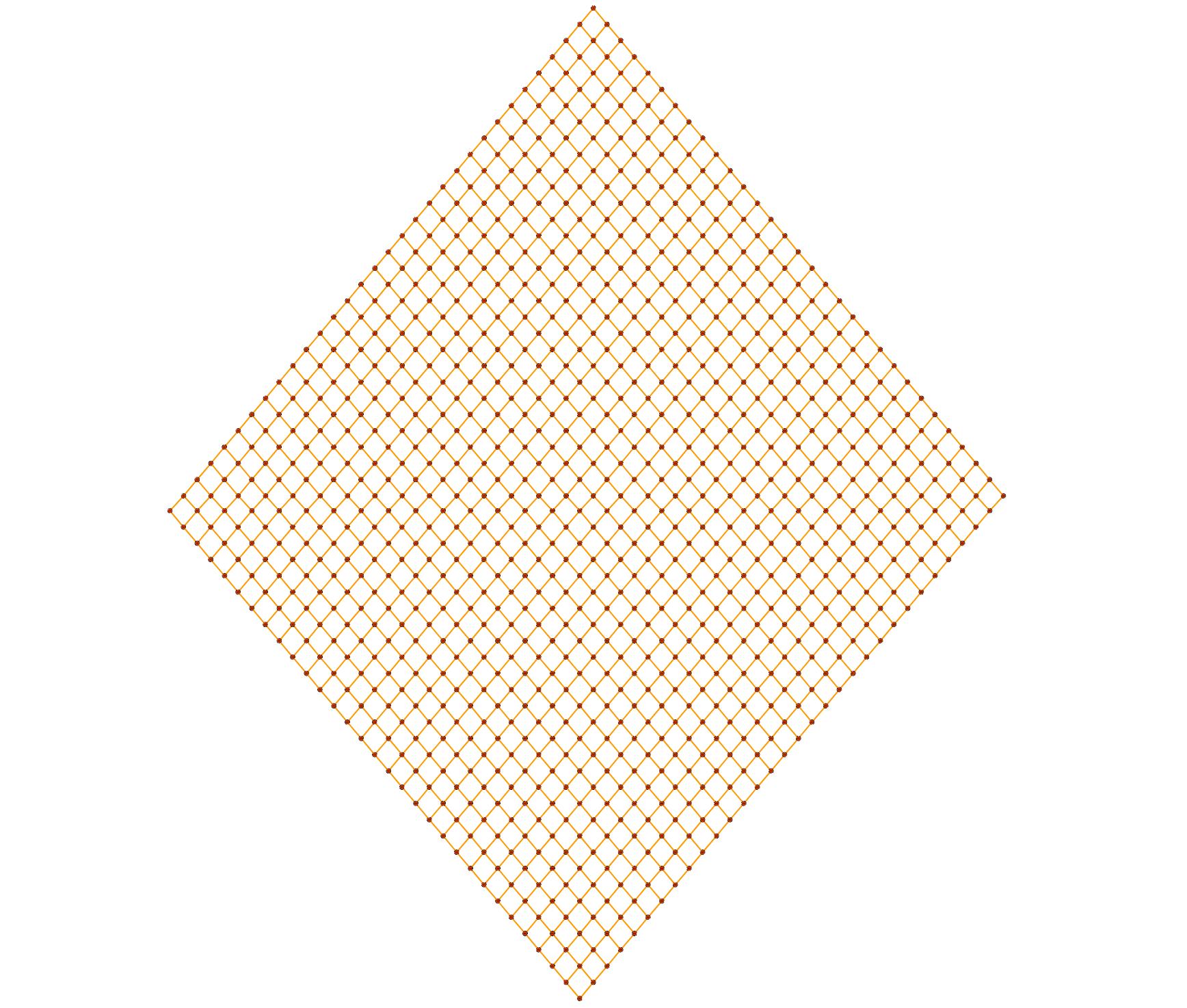}
    \caption[The near-regular network representation]{The n-r 1 network with $N = 992$ and $\bar{k}=3.87$}
    \label{fig:n-r1}      
\end{figure}

The near-regular (n-r) networks are best visualized in a
planar, grid-like fashion. The n-r 1 network is composed of a 31 by 32
grid where node $i$ is connected to node $j$ if $j$ is a distance $d=1$ unit: $1$ unit is the regular distance among nodes in the grid.  The structure of n-r 2 is similar to that of the regular.  However, in addition to $d= 1$ unit, all nodes within a distance of $d=\sqrt{2} units$ are connected. Figure \ref{fig:n-r1} shows the n-r 1 network.

\subsection{Trade-off and Optimization Models}
\begin{figure}[h]
\centering  
    \includegraphics[height=2in]{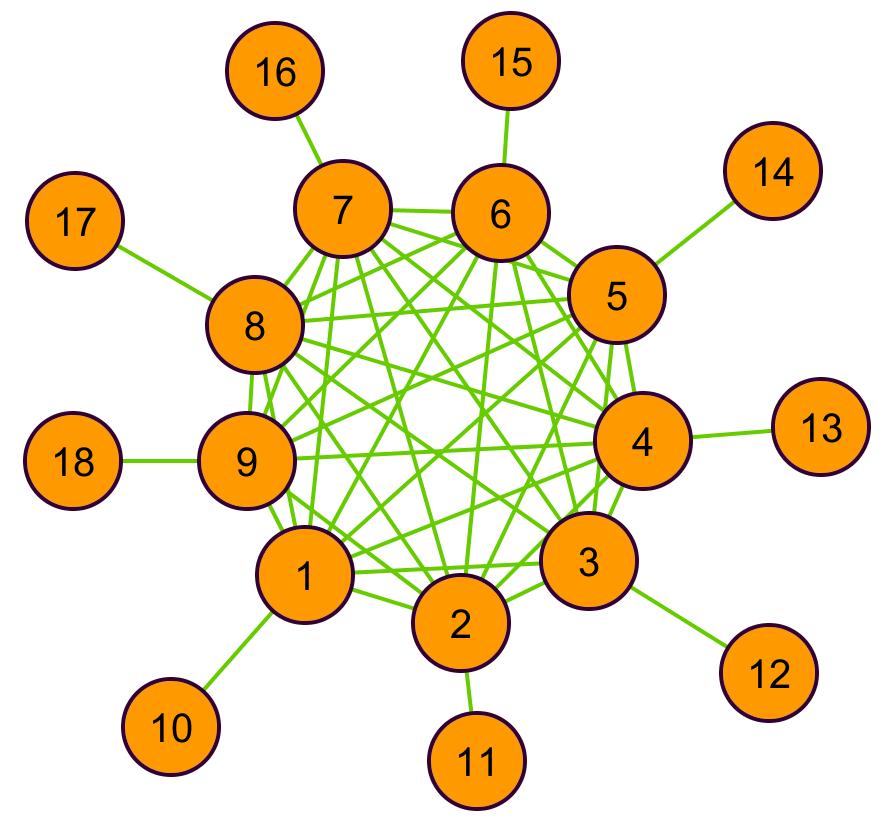}
    \caption[Meshcore network]{The meshcore network with $N=1000$ and $\bar{k}=2.55$}
    \label{fig:meshcore}      
\end{figure}

\begin{figure}[h]
\centering  
    \includegraphics[height=2in]{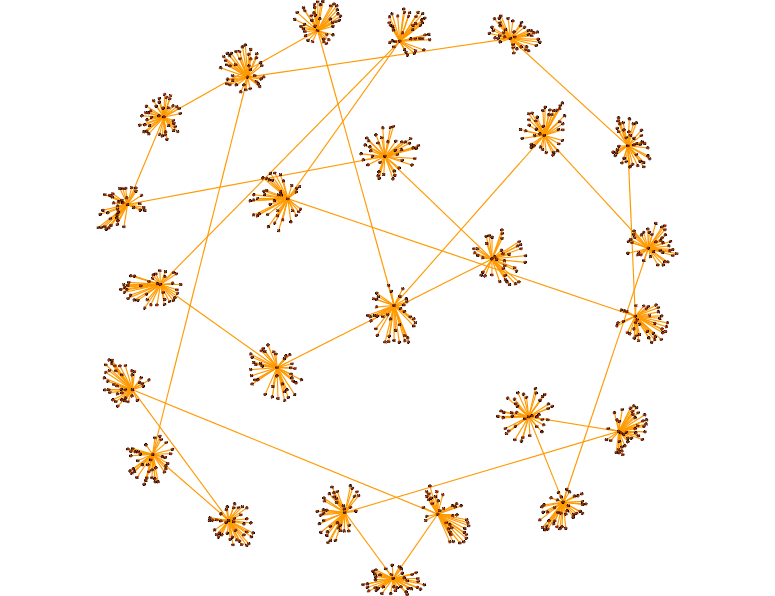}
    \caption[Ringcore network]{The ringcore network with $N=1000$ and $\bar{k}=2$}
    \label{fig:ringcore}      
\end{figure}

\begin{figure}[h]
\centering  
    \includegraphics[height=2in]{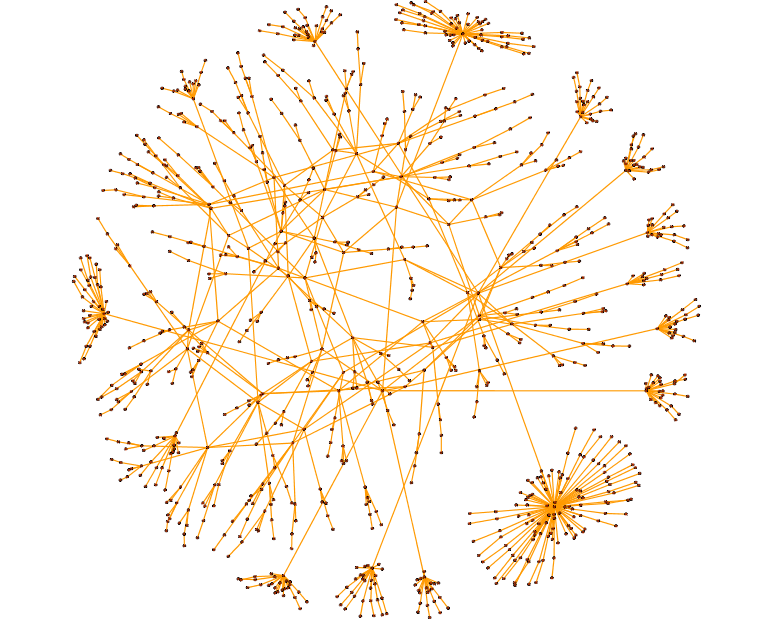}
    \caption[The HOT 2 network]{The HOT 2 network with $N=1000$ and $\bar{k}=2.098$}
    \label{fig:HOT2}      
\end{figure}

The authors of \cite{TGRSHE05} introduce networks with bimodal degree distributions optimized to minimize the impact of random attacks. The meshcore and ringcore topologies shown in Figures \ref{fig:meshcore} and \ref{fig:ringcore}  represent this model. The Heuristically Optimized Trade-off (HOT) network presents a simple model for Internet growth \cite{AECH02,dlwc05}.  The HOT 1 and 2 networks represent this model. Figure \ref{fig:HOT2} shows the HOT 2 network.

\subsection{Real-World Models}
\begin{figure}[h]
\centering  
    \includegraphics[height=2in]{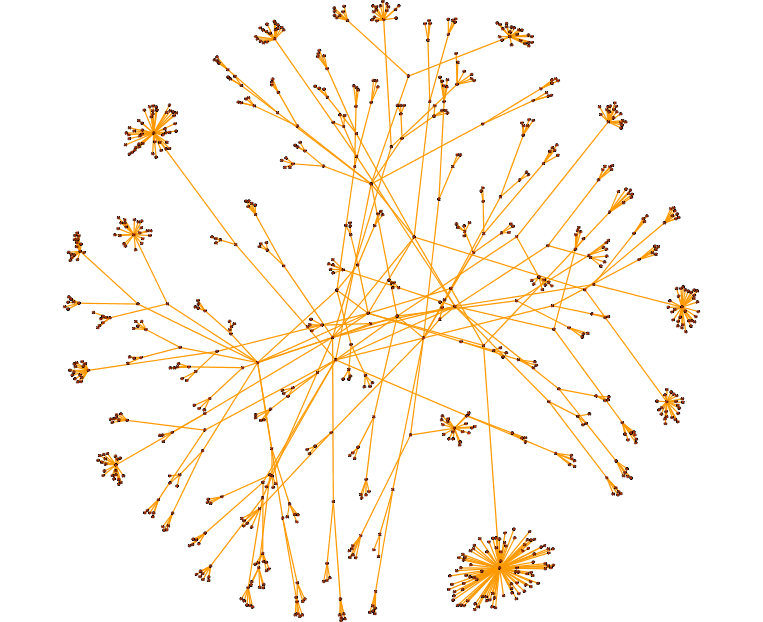}
    \caption[The Abilene network]{The Abilene network with $N=1000$ and $\bar{k}=2.022$}
    \label{fig:abilene}      
\end{figure}

Online social networking connects individuals with common interests. This paper features the MySpace, YouTube, and Flickr networks. These networks were obtained via snowball sampling and have been rescaled \cite{PCDBA07}. The Abilene network in Figure \ref{fig:abilene} was built using the Abilene core while customers and peer networks were each replaced with a gateway router \cite{dlwc05}.

\section{Robustness Metric}
\label{robustness}
The study of robustness is fundamental to numerous network research problems using approaches
that amplify internal behaviors of a network. To this end, we use elasticity as our measure of robustness, obtain its upper bound and finally, select the most feasible routing algorithm for elasticity.

\subsection{Elasticity}
\begin{figure}
\centering  
    \includegraphics[height=2in]{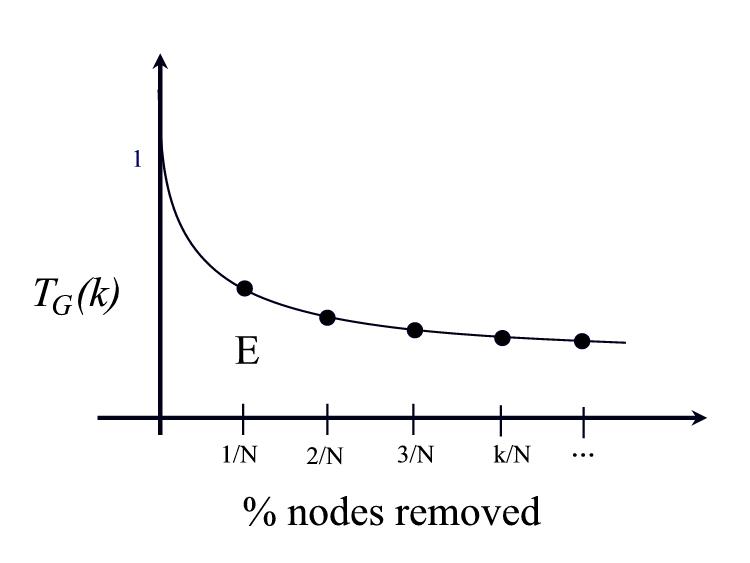}
    \caption[The evaluation of elasticity]{The evaluation of elasticity}
    \label{fig:upperbound}      
\end{figure}

For a network $G$, having no loops or parallel links, elasticity $E(G)$ is a measure of the overall robustness. As shown in Figure \ref{fig:upperbound}, elasticity is the area under the curve of throughput versus the percentage of nodes removed. The throughput is normalized to compare networks of different magnitudes and at each iteration, it is recalculated at the removal of each node.  Initially, $T_{G}\left(0\right) = 1$ which accounts for the normalized throughput.  This value decreases as $\frac{k}{N} \%$ of nodes are removed and therefore, elasticity (E) provides a measure of robustness at any point of node removal.

Therefore, when $\zeta$ nodes have been removed, elasticity can be computed as

\begin{equation}
	\label{eqn:Elasticity}
	E\left( \frac{\zeta}{N} \right)= \frac{1}{2N}\sum_{k=0}^{\zeta} {\left(   T_{G}\left(\frac{k}{N}\right) + T_{G}\left(\frac{k+1}{N}\right)                  \right)   }	
\end{equation}

\begin{flushleft}
where $T_{G}(\frac{k}{N})$ is the throughput at each interval when $k$ nodes are removed.  $N$ is the total number of nodes in the network and $0 \leq \left(\ zeta, k \right) \leq N$.
At each iteration, the throughput is computed as

\begin{equation}
\centering
\label{eqn:Throughput}
  T_{G}\left(t \right)=  \frac{ \max_{\rho}  \sum_{i,j}X_{i,j}\left(t\right) }{\alpha} \hspace{5mm} s.t.\hspace{3mm} LX \leq B\left(t\right)
\end{equation}

 where $t=\frac{k}{N}$ and $\rho$ is a constant used to vary the proportion of flows in network. $\alpha$ is the unnormalized initial throughput and $X_{i,j}\left(t\right)$ is the traffic flow between source node $i$ and destination node $j$.
 $L$ is the routing matrix, $X$ is a vector of all $X_{i,j}\left(t\right)$ flows, and $B\left(t\right) $ is a vector of all link bandwidth capacities.
\end{flushleft}

\subsection{Upper bound for Elasticity}
\label{upperb}
\subsubsection{Analytical results}
In this section, we consider the mesh network as the topology which provides the highest elasticity under all attack strategies for any given network.  We assume homogeneous flows where each flow has a value of $1$.  Additionally, each link has a capacity of $1$ and $X_{ij}\left(t\right)$ can be $1$ or $0$ depending on whether or not a flow exists between nodes $i$ and $j$.  With these assumptions, we proceed to determine the upper bound for elasticity. 

\vspace{.1in}

{\bf Theorem.} Given a mesh network with $N$ nodes, and assuming homogeneous flows and link capacities of $1$, then $\lim_{N\rightarrow\infty}  E(N)= \frac{1}{3}$.

\vspace{.1in}

\begin{flushleft}
\emph{Proof.} Elasticity can be formulated using both discrete and continuous approaches. At each iteration when a node is removed, the throughput is given by

\begin{equation}
\centering
\label{eqn:Through}
  T_{G}\left(t \right)= \frac{\left(N-k\right)\left(N-k-1\right)}{N\left(N-1\right)}
\end{equation}

where $t=\frac{k}{N}$.  
\end{flushleft}

\vspace{.1in}

\emph {Discrete Elasticity (trapezoidal integration).}
For a given network of size $N$, Equation \ref{eqn:Elasticity for n in the discrete case} computes elasticity when $\zeta$ nodes have been attacked. 

\begin{equation}
\centering
\label{eqn:Elasticity for n in the discrete case}
  E\left(\zeta\right)=\frac{1}{N}\left(\frac{1}{2}+ \sum_{k=1}^{\zeta-1} \beta + \delta\right)							           
\end{equation}

\begin{flushleft}
where  $ \beta = \frac{\left(N-k\right)\left(N-k-1\right)}{N\left(N-1\right)} $, $\delta =\frac{\left(N-\zeta\right)\left(N-\zeta-1\right)}{2N\left(N-1\right)} $, and $\zeta \leq N-1$.
Equation \ref{eqn:Elasticity for N in the discrete case} computes the total elasticity for a network with $N$ nodes when all $N$ nodes are progressively removed. 
\end{flushleft}

\begin{equation}
\centering
\label{eqn:Elasticity for N in the discrete case}
  E\left(N\right)=\frac{1}{N}\left(\frac{1}{2}+ \sum_{k=1}^{\zeta-1} \frac{\left(N-k\right)\left(N-k-1\right)}{N\left(N-1\right)}\right) 
\end{equation}

\emph{ Continuous Elasticity}
Equation \ref{eqn:econt} gives the formulation of elasticity for the continuous case. Similar to the discrete case, Equation \ref{eqn:Elasticity for n in the continuous case} computes elasticity for a given mesh network with size of $N$ where $\zeta$ nodes have been removed and Equation \ref{eqn:Elasticity for N in the continuous case} computes the total elasticity for a mesh network with $N$ nodes.  As the size of the network grows, Equation \ref{eqn:limit of elasticity} then provides the upper bound on elasticity when all $N$ nodes are removed.

\begin{equation}
\centering
\label{eqn:econt}
  E\left(t\right)=\int_{0}^{t} {T_{G}\left(\tau \right) d \tau},\hspace{.1in} 0 \leq t \leq 1
\end{equation}

\begin{equation}
\centering
\label{eqn:Elasticity for n in the continuous case}
  E\left(\zeta\right)=\frac{N\left(N-1\right)\zeta + \frac{1}{2}\left(1-2\zeta\right)\zeta^{2} + \frac{1}{3}\zeta^{3} }{ N^{2}\left(N-1\right)}
\end{equation}

\begin{equation}
\centering
\label{eqn:Elasticity for N in the continuous case}
  E\left(N\right)=\frac{1}{3} - \frac{1}{6N} - \frac{1}{6N^2} 
\end{equation}

\begin{flushleft}
Therefore,
\end{flushleft}

\begin{equation}
\centering
\label{eqn:limit of elasticity}
  \lim_{N\rightarrow\infty}  E(N)= \frac{1}{3}
\end{equation}

\begin{flushleft}
Q.E.D.
\end{flushleft}

\subsubsection{Numerical Results}
\begin{figure}[h]
	\centering
		\includegraphics[height=2in]{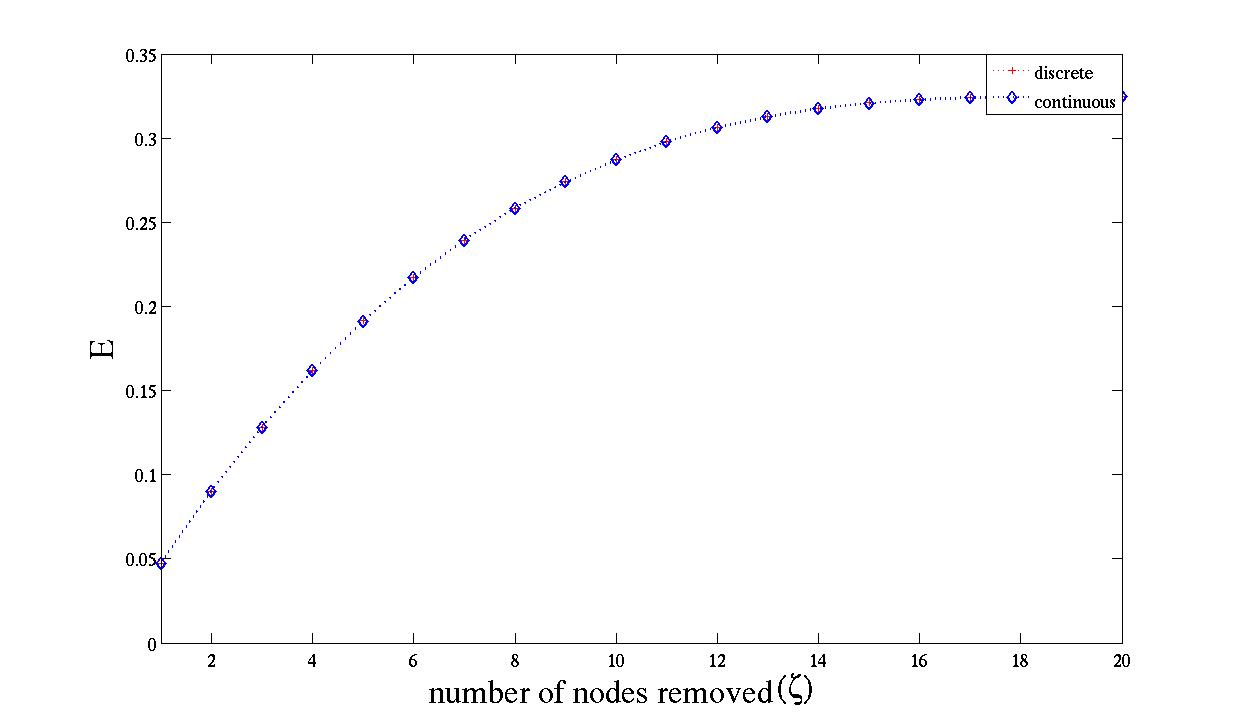}
		 \caption[Convergence of elasticity]{Comparison of the convergence rates of elasticity, from Equations \ref{eqn:Elasticity for n in the discrete case} and \ref{eqn:Elasticity for n in the continuous case}, where $\zeta$ nodes have been attacked. }
	\label{fig:discrete_cont_smalln}
\end{figure}

Figure \ref{fig:discrete_cont_smalln} compares the convergence rate of the discrete and continuous cases when $\zeta$ nodes have been attacked from a network where $N=20$. As depicted, both approaches converge at the onset of node removal.

\begin{figure}[h]
	\centering
		\includegraphics[height=1.9in]{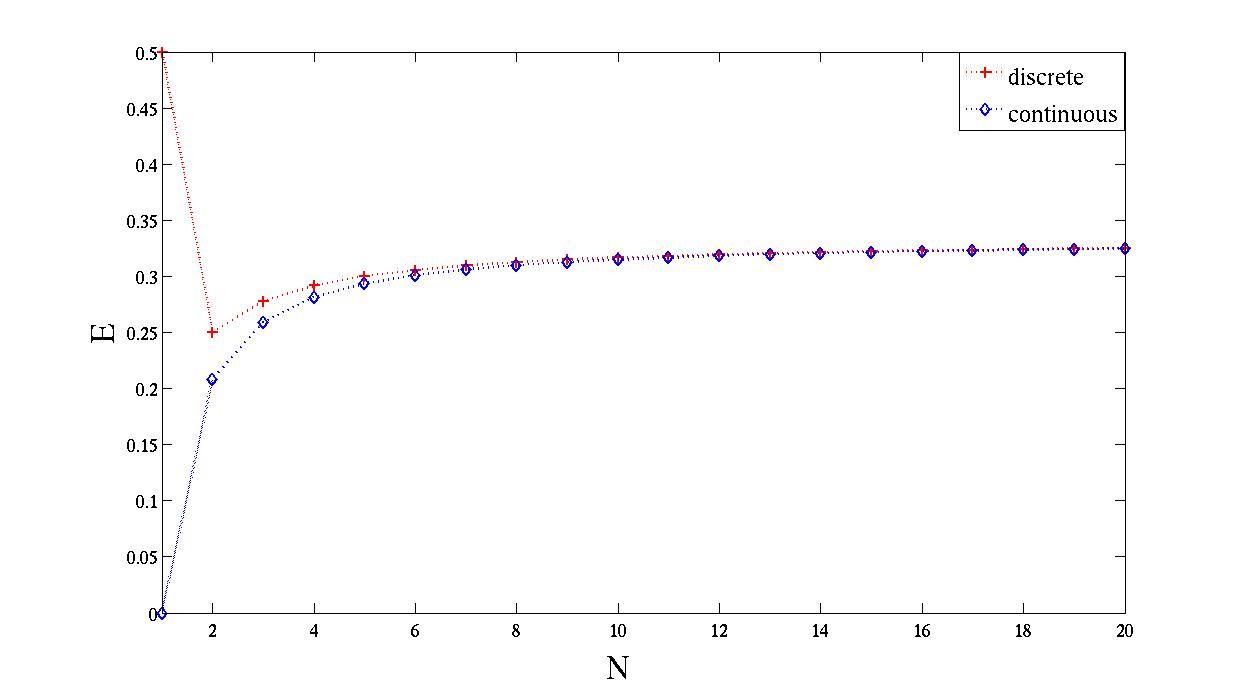}
		 \caption[Convergence of elasticity]{Comparison of the convergence rates of elasticity, from Equations \ref{eqn:Elasticity for N in the discrete case} and \ref{eqn:Elasticity for N in the continuous case}, for a network of size $N$}
	\label{fig:discrete_cont_N}
\end{figure}

Figure \ref{fig:discrete_cont_N} compares the convergence rate of elasticity for the discrete case in Equation \ref{eqn:Elasticity for N in the discrete case} to the continuous case in Equation \ref{eqn:Elasticity for N in the continuous case}. As shown, both cases converge for a network with 10 nodes.    

These convergence rates are significant because they necessitate few iterations. More importantly, the discrete approaches can be abandoned for the continuous approaches to simplify calculations without compromising accuracy.

\subsection{Routing Algorithm for Elasticity}
\label{alg_mod}

Elasticity depends on the routing algorithm selected.  For this reason, three routing approaches are explored: 1) Optimization (heterogeneous traffic matrix), 2) Dijkstra's Algorithm (heterogeneous traffic matrix), and 3) Dijkstra's Algorithm (homogeneous traffic matrix). All approaches assumed homogeneous link capacities of $1$.

\subsubsection{Optimization (Heterogeneous traffic matrix)}

The Objective Function (Function \ref{eqn:objective}) of the optimization problem maximizes the individual flow between any pair of nodes. Equations \ref{eqn:constr1}-\ref{eqn:constr4} are the main constraints to the optimization problem. Equation \ref{eqn:constr1} ensures that each node sends $\delta$ unit of traffic to every node, while Equation \ref{eqn:constr2} represents the balance of the incoming and outgoing traffic demands through any node in the network. Inequality \ref{eqn:constr3} represents the capacity constraint on each link, and Equation \ref{eqn:constr4} computes the utilization of each link.

\begin{equation}
\label{eqn:objective}
Maximize \ \delta
\end{equation}

\text{Subject to}

\begin{equation}
\label{eqn:constr1}
\sum_{j \in N} flow_{s,j,s} = \delta (N-1) \ \forall s
\end{equation}

\begin{equation}
\label{eqn:constr2}
\sum_{i \in N} (flow_{i,j,s} - flow_{j,i,s}) = \delta \ \forall s, j, j \neq s
\end{equation}

\begin{equation}
\label{eqn:constr3}
\sum_{s \in N} flow_{i,j,s} \leq capacity_{i,j} \ \forall i, j, i \neq j
\end{equation}

\begin{equation}
\label{eqn:constr4}
utilization_{i,j} = \sum_{s \in N} flow_{i,j,s} \ \forall i, j
\end{equation}

\vspace{.1in}
Algorithm \ref{alg:hetero} provides elasticity using the optimization approach discussed previously.

\begin{algorithm}
\caption{Optimization}
\label{alg:hetero}
\begin{algorithmic}

\WHILE {$Connected:=True$}
\STATE $capacity_{i,j}:=1$\\
\STATE $demand_{i,j}:=0$\\

\WHILE {$\sum_{i,j} capacity_{i,j} \neq 0$} 
\STATE Solve the optimization problem \\
\STATE Update the demand between nodes that are connected with non-zero capacity links\\
\STATE $demand_{i,j}:=demand_{i,j}+\delta$\\
\STATE $capacity_{i,j} := capacity_{i,j} - utilization_{i,j}$\\
\ENDWHILE

\STATE Remove one node (or a group of nodes)\\
\ENDWHILE

\end{algorithmic}
\end{algorithm}

\subsubsection{Dijkstra's algorithm (heterogeneous traffic matrix)}
The second approach realizes Dijkstra's algorithm. As shown in Algorithm \ref{alg:dijalg}, flows traverse the shortest path from source to destination. This algorithm has a running time $O(n^{2})$.  However, when the heterogeneous traffic matrix is considered, the running time increases to $O(n^{3})$.

\begin{algorithm}
\caption{Dijkstra's algorithm}
\label{alg:dijalg}
\begin{algorithmic}
\STATE \emph{\bf begin}\\
\STATE $S:=0; \bar{S}:=N$ 
\STATE $d(i):=\infty$ for each node $i \in N$
\STATE $d(s):= 0$ and pred(s) $:=0$
\STATE \emph{\bf while} $\left|S\right|<n$ \emph{\bf do}\\
\STATE \emph{\bf begin}\\
\STATE let $i \in \bar{S}$ be a node for which $d\left(i\right)=\min \left\{d\left(j\right) : j \in \bar{s} \right\}$\\
\STATE $S:=S\cup\left\{i\right\};$\\
\STATE $\bar{S}:=\bar{S}-\left\{i\right\};$\\
\FOR {each $\left(i,j\right) \in A\left(i\right)$}
   \STATE \emph{\bf if} $d\left(j\right)>d\left(i\right)+ c_{ij}$ \emph{\bf then} $d\left(j\right) := d\left(j\right) + c_{ij}$ and pred($j$) $:=i$
 \ENDFOR  
   \STATE \emph{\bf end} 

\end{algorithmic}
\end{algorithm}

\subsubsection{Dijkstra's algorithm (homogeneous traffic matrix)}
\label{approach3}
This approach also revolves around Algorithm \ref{alg:dijalg} and likewise, has a running time $O(n^{2})$.  However, a homogeneous traffic matrix was implemented. Given these three models, Subsection \ref{evalmods} evaluates each and selects the most feasible.

\subsection{Evaluation of Routing Models}
\label{evalmods}

\begin{figure}[h]
\centering  
    \includegraphics[height=1.25in]{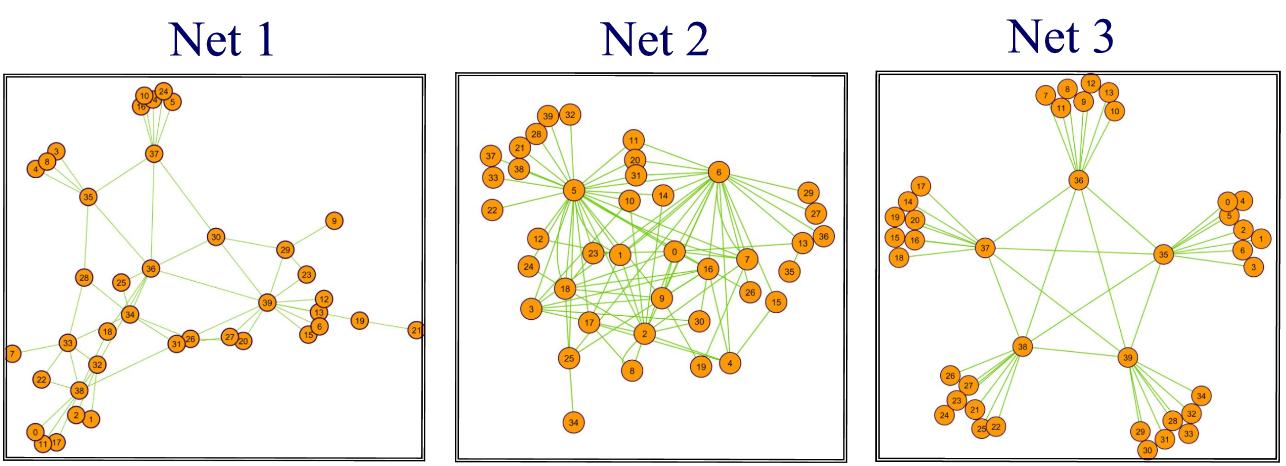}
    \caption[Hot topology]{Three networks for which elasticity was evaluated}
    \label{fig:hothetop}      
\end{figure}

Figure \ref{fig:hothetop} shows the three networks for which elasticity was computed: Net 1, Net 2, and Net 3. For these three networks, we compare the results of elasticity provided by each routing algorithm targeting first, nodes with the highest degree and second, nodes with highest betweenness.

\begin{figure}[htbp]
\centering  
    \includegraphics[height=2in]{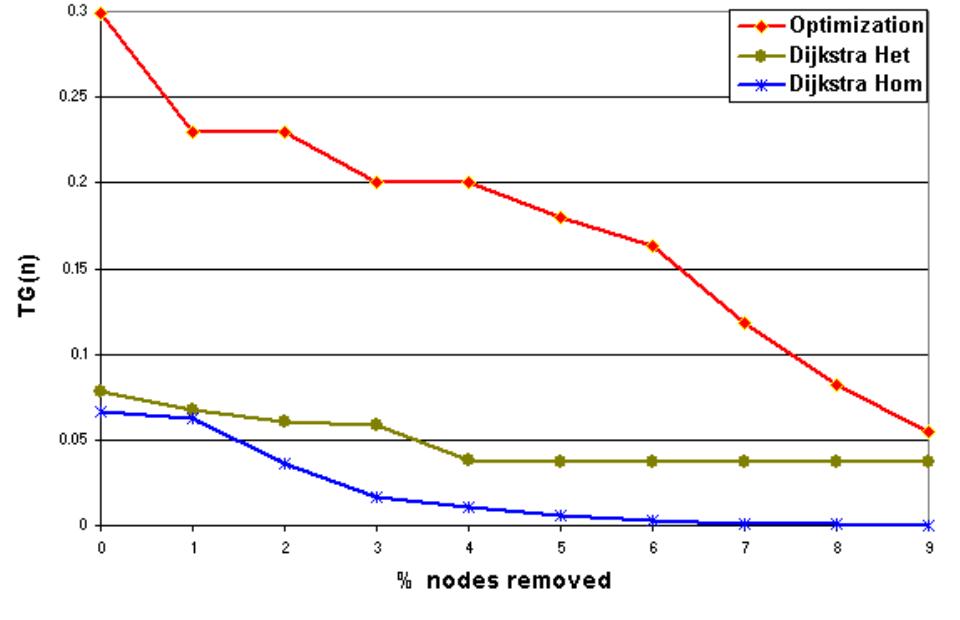}
    \caption[Hot topology]{Throughput degradation as nodes with highest degree are attacked for Net 1. ``Het" represents a heterogeneous traffic matrix and ``Hom" represents homogeneous traffic matrix.}
    \label{fig:hothetopgra}      
\end{figure}

Figure \ref{fig:hothetopgra} shows the throughput degradation as nodes with highest degree are attacked in Net 1.  As depicted, the optimization approach produces the highest elasticity, followed by Dijkstra's heterogeneous approach and finally, Dijkstra's homogeneous approach.  This trend was observed for each network under both attack strategies.  However, under certain circumstances where the network has low connectivity, the elasticity results were identical for both Dijkstra's ``heterogeneous" and optimization approach.

For each of the three routing approaches, each network was given a rank of 1, 2 or 3, based on its value for elasticity: 1 as the highest and 3 as the lowest.  Table \ref{tab:HNDa} displays the rankings for each network under highest node degree attack.  As shown, elasticity was highest for Net 1, followed by Net 2, and finally, Net 3, for each approach.  Though the values were different for the highest betweenness attack strategy (not shown), the rankings were similar to that of Table \ref{tab:HNDa}.  

\begin{table}[h]
 \centering
  \caption{Elasticity comparison for all networks under {\bf highest node degree} attack}
    \begin{tabular}[c]{|c|c|c|c|}
        \hline
         Algorithm     & Rank 1 & Rank 2 & Rank 3 \\
        \hline
          1   &        Net 1  & Net 2 & Net 3  \\ \hline 
          2   &        Net 1  & Net 2 & Net 3  \\ \hline     
          3   &        Net 1  & Net 2 & Net 3 \\ \hline  
    \end{tabular}   
    \label{tab:HNDa}  
\end{table}

Furthermore, we observed that the criteria for node addition to the shortest path could potentially affect the results of elasticity.  More specifically, in Algorithm \ref{alg:dijalg}, nodes are added to the shortest path if the following optimality condition is satisfied:
\begin{equation}
d\left(j\right)>d\left(i\right)+ c_{ij}	
\end{equation}

\begin{flushleft}
where $d(j)$ is the distance label at node $j$ and $c_{ij}$ is the cost of moving from node $i$ to $j$.  
\end{flushleft}
 
However, if there are several nodes $j$, such that each node satisfies this condition, the next node added to the shortest path is selected sequentially. To investigate the impact of this constraint on elasticity, we modify Algorithm \ref{alg:dijalg} to relax the sequential constraint by randomly selecting the next node $j$ that will be added to the shortest path. Algorithm \ref{alg:dijalgmod} reflects these changes.  

\begin{algorithm}
\caption{Dijkstra's ``Modified" algorithm}
\label{alg:dijalgmod}
\begin{algorithmic}
\STATE \emph{\bf begin}\\
\STATE $S:=0; \bar{S}:=N; X=0$ 
\STATE $d(i):=\infty$ for each node $i \in N$
\STATE $d(s):= 0$ and pred(s) $:=0$
\STATE \emph{\bf while} $\left|S\right|<n$ \emph{\bf do}\\
\STATE \emph{\bf begin}\\
\STATE let $i \in \bar{S}$ be a node for which $d\left(i\right)=\min \left\{d\left(j\right) : j \in \bar{s} \right\}$\\
\STATE $S:=S\cup\left\{i\right\};$\\
\STATE $\bar{S}:=\bar{S}-\left\{i\right\};$\\
\FOR {each $\left(i,j\right) \in A\left(i\right)$}
   \STATE $X_{i} = j, \forall j \in N$ which satisfy the optimality condition\\
   \STATE $j_{selected}=rand\left(X_{i}\right)$\\
    \STATE \emph{\bf then} $d\left(j\right) := d\left(j\right) + c_{ij}$ and pred($j$) $:=i$ \\ 
   \ENDFOR
   \STATE \emph{\bf end}

\end{algorithmic}
\end{algorithm}

We conducted 100 sample runs and averaged elasticity for each network under highest degree and highest betweenness attacks.  Our results show a negligible difference between the elasticity results for Algorithm \ref{alg:dijalg} and \ref{alg:dijalgmod}.  Hence, the rankings shown in Table \ref{tab:HNDa} remain the same.  

From the three algorithms, we select Dijkstra's algorithm, using a homogeneous traffic matrix, as the most feasible because it produces qualitatively comparable results to the other two algorithms and has the least costly running time: $O(n^{2})$. 

\section{Experimental Results}
\label{experimentalresults}

In this Section, we evaluate elasticity for a set of selected topologies.  First, we compute the elasticity of all networks under each attack strategy and second, we implement a tradeoff function that combines the elasticity obtained for each attack strategy and penalizes networks for having excess links. 

\subsection{Elasticity of Networks Under Three Attack Strategies}
In the subsequent sections, Elasticity R, Elasticity D, and Elasticity B refer to elasticity under the following three attack strategies:

\vspace{.1in}

\begin{enumerate}
	\item removal of random nodes (Elasticity R)
	\item removal of highest degree nodes (Elasticity D)
	\item removal of highest betweenness nodes (Elasticity B) 
\end{enumerate}

\vspace{.1in} 

Table \ref{tab:attacks} ranks all networks in descending order of magnitude based on the number of links and the scores for elasticity under the three strategies. As shown, the mesh network is the most robust under all strategies.  This is expected, as it sets the upper bound on elasticity. Under random attacks, the elasticity for the Gi-dense and MySpace networks are in proximity to that of the mesh network.  As cost is a critical factor in network design, it is financially sensible to implement the latter two topologies rather than the mesh because Table \ref{tab:attacks} shows that the MySpace and Gi-dense networks can provide about $94\%$ of the elasticity that the mesh provides while only using about $1\%$ of the links. 

\begin{table}[h]
\centering
\caption{Networks sorted in descending order for number of links, Elasticity R (Elas. R), Elasticity D (Elas. D), and Elasticity B (Elas. B)} 
    \begin{tabular}[l]{|c|c||c|c||c|c||c|c|}
           \hline
        {\bf Nets.}   &  { \bf links} &  {\bf Nets.}    &  { \bf Elas. R} &  {\bf Nets.}    &  { \bf Elas. D}  &  { \bf Nets.} & { \bf Elas. B}  \\
        \hline
		mesh	&	499500	&	mesh	&	0.3333	&	mesh	&	0.3333	&	mesh	&	0.3333	\\ \hline
		MySpace	&	10976	&	MySpace	&	0.3119	&	n-r 2	&	0.2426	&	Gi-dense	&	0.2390	\\ \hline
		Gi-dense	&	4505	&	Gi-dense	&	0.3111	&	Gi-dense	&	0.2082	&	W-S 2	&	0.1770	\\ \hline
		n-r 2	&	3781	&	PA 2	&	0.2743	&	MySpace	&	0.1721	&	MySpace	&	0.1719	\\ \hline
		W-S 2	&	3000	&	W-S 2	&	0.2703	&	W-S 2	&	0.1640	&	W-S 1	&	0.1260	\\ \hline
		PA 2	&	2964	&	PA 1	&	0.2677	&	n-r 1	&	0.1342	&	Gi-sparse	&	0.1010	\\ \hline
		Gi-sparse	&	2009	&	Gi-sparse	&	0.2520	&	W-S 1	&	0.1170	&	PA 2	&	0.0719	\\ \hline
		W-S 1	&	2000	&	W-S 1	&	0.2490	&	Gi-sparse	&	0.1143	&	PA 1	&	0.0558	\\ \hline
		PA 1	&	1981	&	n-r 2	&	0.2316	&	PA 2	&	0.0644	&	YouTube	&	0.0332	\\ \hline
		n-r 1	&	1921	&	Flickr	&	0.2211	&	PA 1	&	0.0535	&	Flickr	&	0.0315	\\ \hline
		YouTube	&	1576	&	YouTube	&	0.2132	&	YouTube	&	0.0371	&	n-r 2	&	0.0246	\\ \hline
		Flickr	&	1515	&	meshcore	&	0.1997	&	Flickr	&	0.0285	&	n-r 1	&	0.0178	\\ \hline
		meshcore	&	1275	&	HOT 2	&	0.1623	&	HOT 1	&	0.0129	&	meshcore	&	0.0083	\\ \hline
		HOT 2	&	1049	&	PA-sparse	&	0.1537	&	HOT 2	&	0.0095	&	HOT 1	&	0.0059	\\ \hline
		PA-sparse	&	1049	&	HOT 1	&	0.1405	&	Abilene	&	0.0093	&	HOT 2	&	0.0048	\\ \hline
		ringcore	&	1000	&	ringcore	&	0.1290	&	meshcore	&	0.0083	&	PA-sparse	&	0.0039	\\ \hline
		HOT 1	&	988	&	Abilene	&	0.1280	&	PA-sparse	&	0.0045	&	Abilene	&	0.0031	\\ \hline
		Abilene	&	896	&	n-r 1	&	0.1016	&	ringcore	&	0.0040	&	ringcore	&	0.0026	\\ \hline

    \end{tabular}
    
    \label{tab:attacks}     
\end{table}

The subsequent Subsections show correlations for elasticity under the specified attack strategy.

\subsubsection{Correlation Between Elasticity and Number of Links}

\begin{figure}[h]
	\centering
		\includegraphics[height=2in]{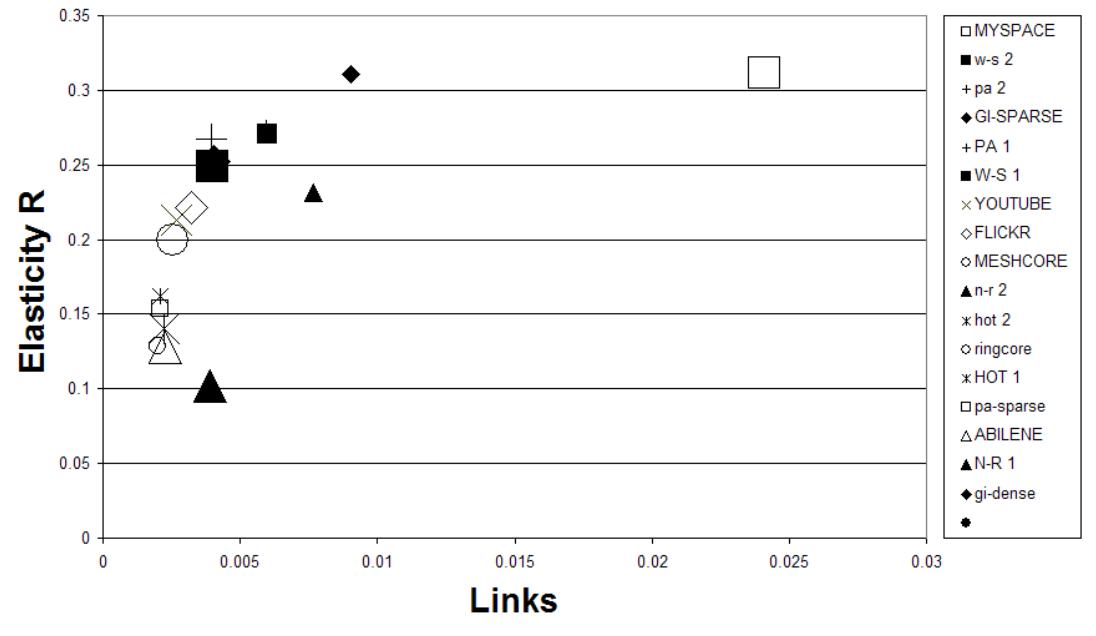}
	 \caption[Elasticity R vs number of links  for each network in Table \ref{tab:attacks}]{Elasticity R vs number of links  for each network in Table \ref{tab:attacks} }
	\label{fig:edgesrand}
\end{figure}

\begin{figure}[h]
	\centering
		\includegraphics[height=2in]{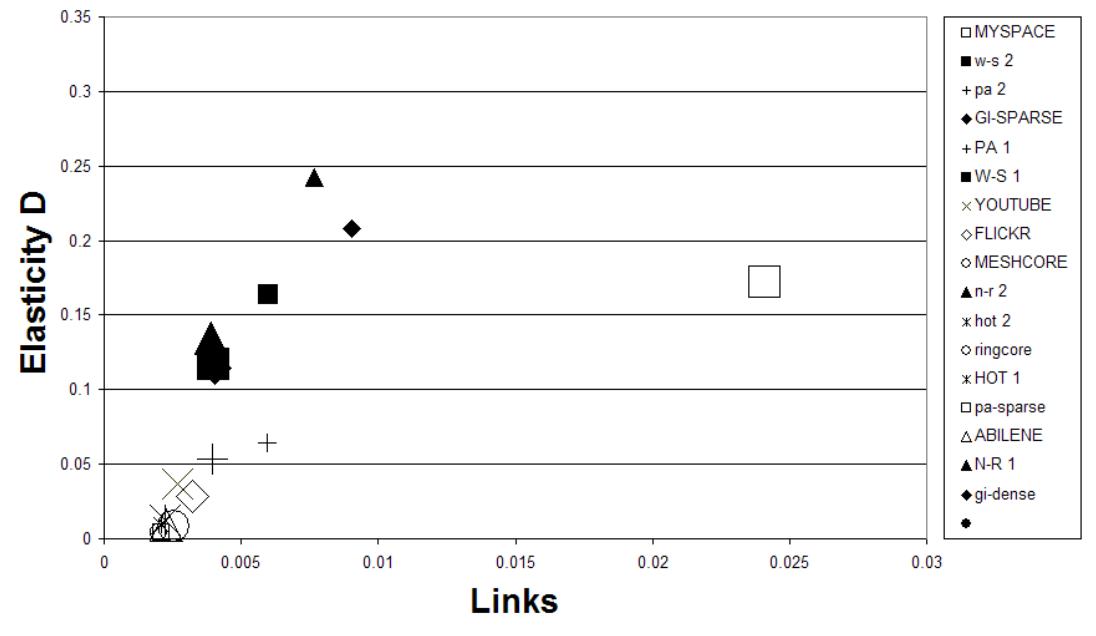}
		 \caption[Elasticity D vs number of links  for each network in Table \ref{tab:attacks}]{Elasticity D vs number of links  for each network in Table \ref{tab:attacks}}
	\label{fig:edgeshigh}
\end{figure}

\begin{figure}[h]
	\centering
		\includegraphics[height=2in]{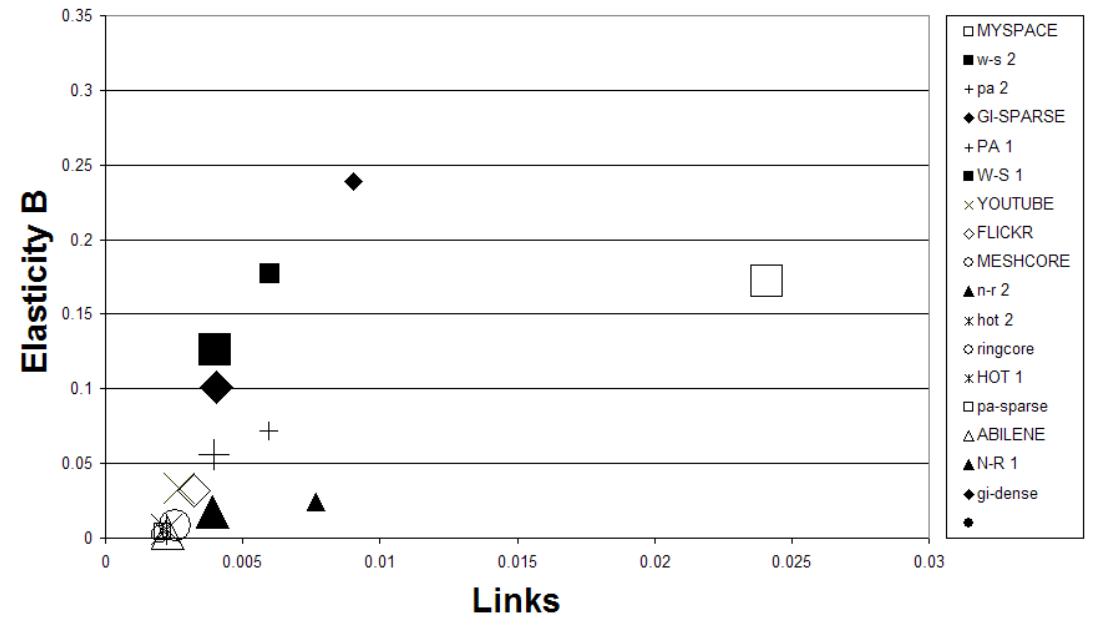}
		 \caption[Elasticity B vs number of links for each network in Table \ref{tab:attacks}]{Elasticity B vs number of links for each network in Table \ref{tab:attacks}}
	\label{fig:edgesbet}
\end{figure}

From Table \ref{tab:attacks} it is notable that for all removal strategies the MySpace, Gi, PA, and Watts-Strogatz networks all vie for the highest elasticity.  This phenomenon can be explained by considering the large number of links of these networks. Figures \ref{fig:edgesrand}, \ref{fig:edgeshigh}, and \ref{fig:edgesbet} confirm this propensity and depict elasticity under random, targeted, and highest betweenness attacks respectively. In these figures, each network is assigned a symbol representative of two classes of networks: 1) The heterogeneous class with graphic or unshaded symbols represents networks with a power-law distribution, and 2) the semi-regular class, further broken down into deterministic and random networks, are the blocked, shaded symbols and is indicative of networks with a Poisson degree distribution. Furthermore, each symbol within a class can be one of two sizes: The large symbols correspond to the networks shown in ``Full caps" in the legend and the small symbols correspond to networks in ``Lower caps." These Figures show that the tendency for elasticity to increase as the number of links increase is not always the case.  Thus, a large number of links is not a necessary condition even if it is a sufficient condition for high elasticity.

Table \ref{tab:attacks} shows that under random attack, elasticity can be as low as $30.5\%$ of the upper bound.  This sharply declines to $1.2\%$ for highest degree attacks and $0.8\%$ for highest betweenness attacks.  For this reason, the design of a robust topology is of utmost importance to obtain high elasticity. For example, the HOT 1 and PA-sparse networks have the same number of links, the same number of nodes, and almost identical degree distributions.  However, their response to attacks differ \cite{dlwc05}. Under random attacks, the PA-sparse provides $9.76\%$ more elasticity than the HOT 1 network. In the PA-sparse network, low degree nodes outnumber high degree nodes (hubs) and hence, the probability of attacking hubs is lower than that of attacking other nodes. This is also the case for the HOT 1 network. However, the ratio of low degree nodes to hubs is higher in the PA-sparse network than in the HOT 1 network.  As a result, Elasticity R for the PA-sparse network is higher than the HOT 1 network.  For highest degree attack the elasticity provided by both networks decreases. However,  the HOT 1 topology provides about three times Elasticity D as the PA-sparse network.  The PA-sparse network is more susceptible to this attack because the hubs in this network facilitate interconnection and are vital to the elasticity of the network. However, the hubs in the HOT 1 network are located on the periphery and are less critical to interconnections \cite{rhal00}.

\begin{figure}[h]
	\centering
		\includegraphics[height=2in]{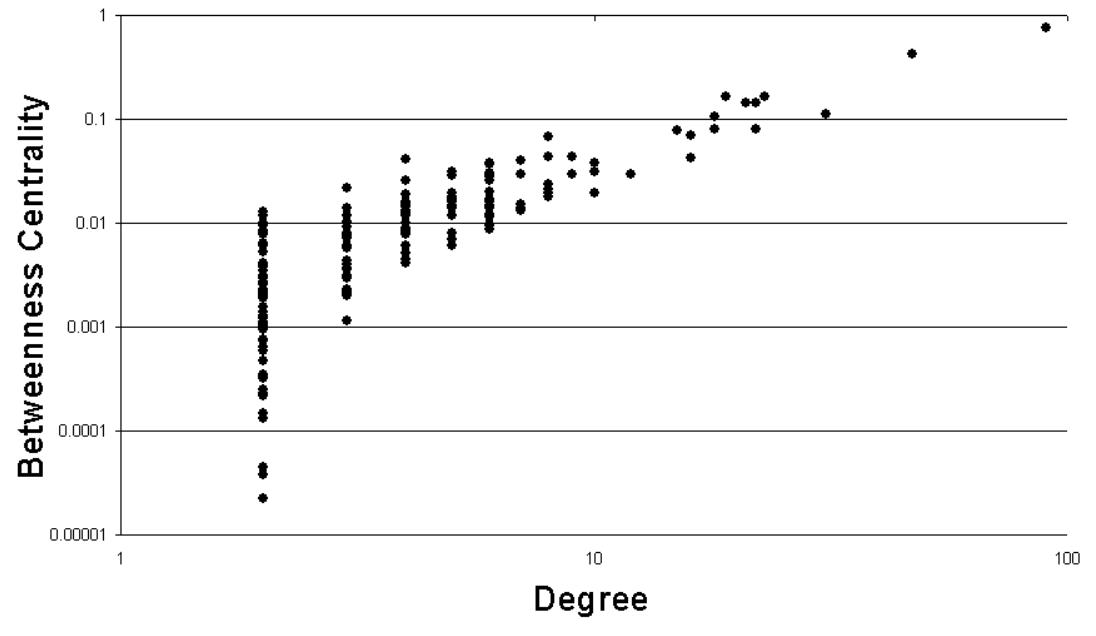}
		 \caption[Convergence of elasticity]{Betweenness distribution for the PA-sparse network}
	\label{fig:betweennessPAsparse}
\end{figure}

\begin{figure}[h]
	\centering
		\includegraphics[height=2in]{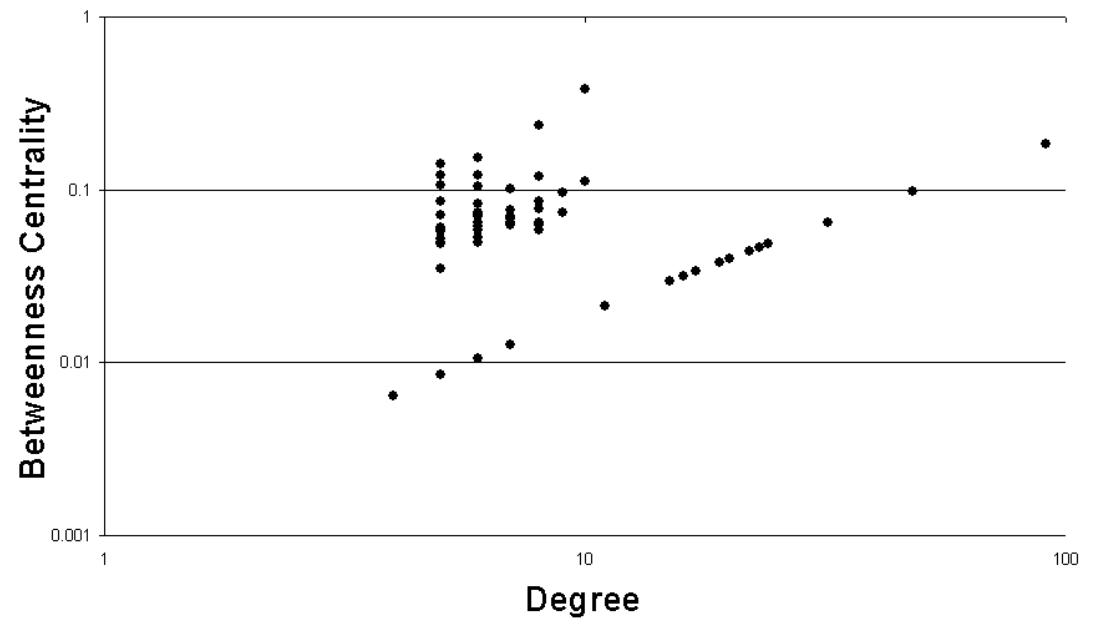}
		 \caption[Convergence of elasticity]{Betweenness distribution for the HOT 1 network}
	\label{fig:betweennessHOT1}
\end{figure}

For highest betweenness attack, the elasticity of both networks decreases even more.  It is notable that from highest degree to highest betweenness attack, the elasticity provided by HOT 1 exhibits a $54.3\%$ decrease whereas that provided by PA 1 exhibits a much smaller decrease of $13.3\%$.  This can be interpreted from Figures \ref{fig:betweennessPAsparse} and \ref{fig:betweennessHOT1} that show the betweenness distribution for the PA-sparse and HOT 1 networks. For the PA-sparse, nodes with the highest degrees have the highest betweenness.  Thus, damage incurred under highest betweenness attacks is almost similar to that under highest degree attack. However, for the HOT 1 network there is a large decrease in elasticity from highest degree attack to highest betweenness attacks because nodes with highest betweenness tend to have lower degrees and facilitate interconnection within the network.  Thus, attacks on these nodes are more detrimental than high degree attacks.

\subsubsection{Correlation Between Elasticity and Heterogeneity}

\begin{figure}[h]
	\centering
		\includegraphics[height=2in]{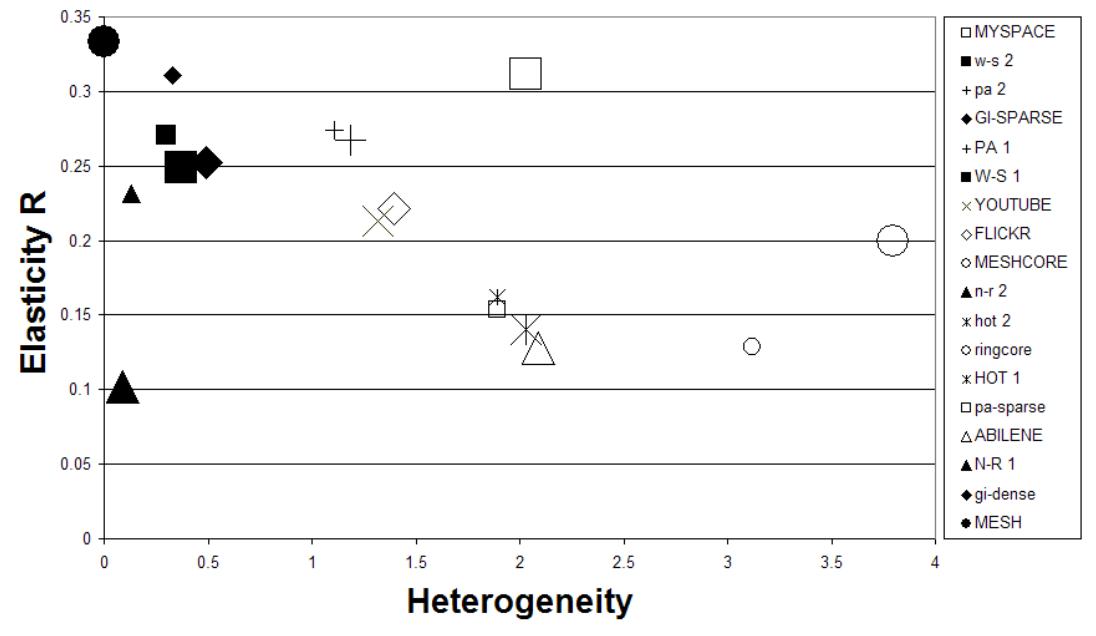}
		 \caption[Convergence of elasticity]{Elasticity R vs heterogeneity  for each network in Table \ref{tab:attacks} }
	\label{fig:hetrand}
\end{figure}

\begin{figure}[h]
	\centering
		\includegraphics[height=2in]{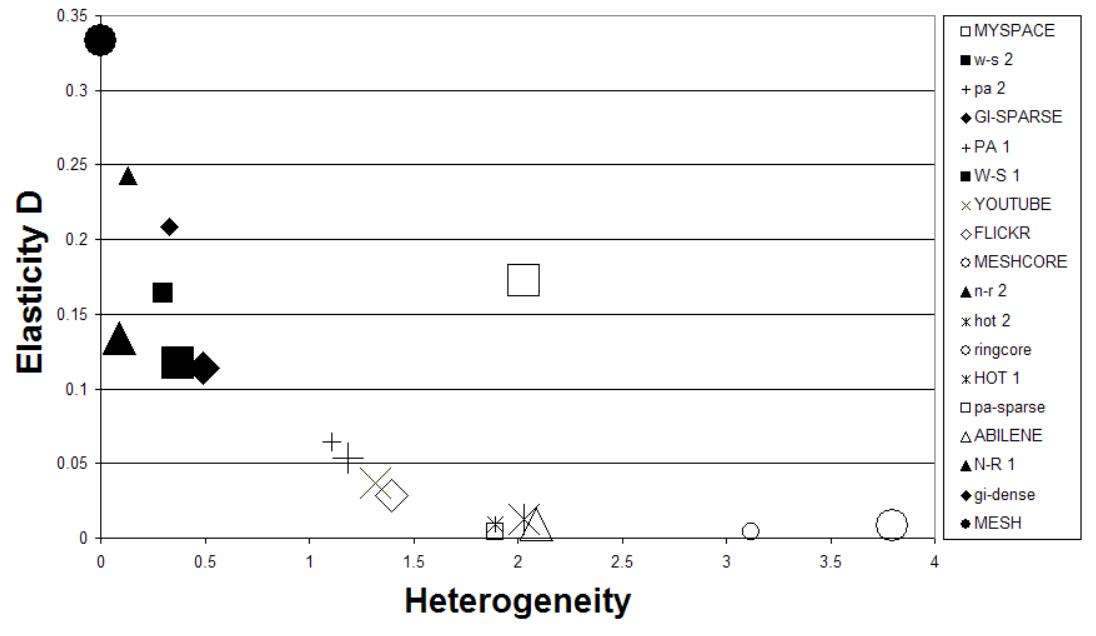}
		 \caption[Convergence of elasticity]{Elasticity D vs heterogeneity  for each network in Table \ref{tab:attacks}}
	\label{fig:hethigh}
\end{figure}

\begin{figure}[h]
	\centering
		\includegraphics[height=2in]{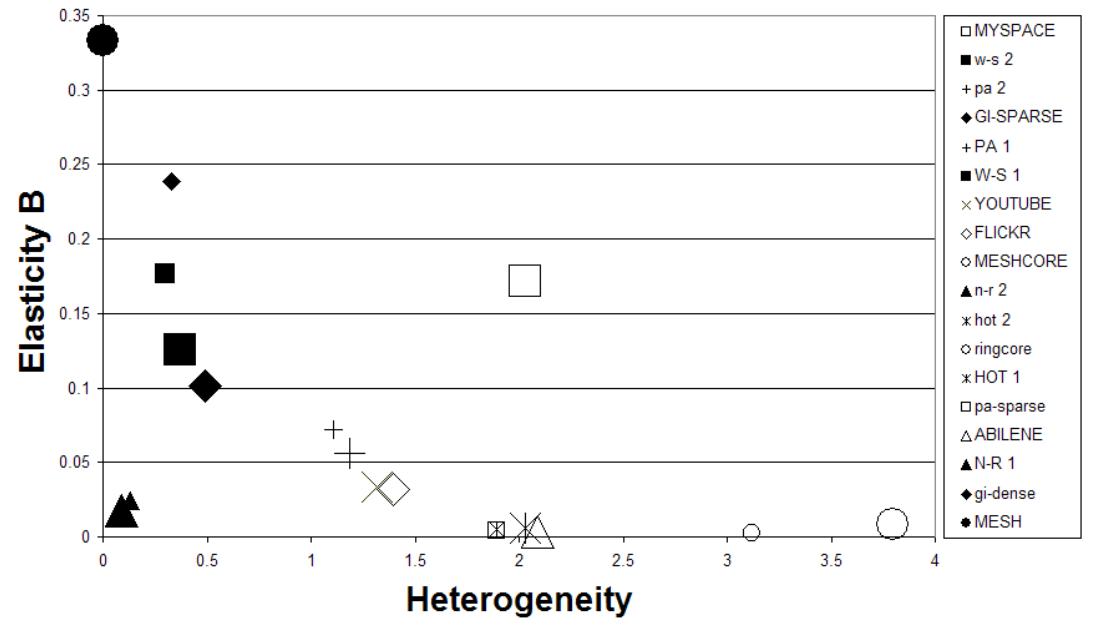}
		 \caption[Convergence of elasticity]{Elasticity B vs heterogeneity  for each network in Table \ref{tab:attacks}}
	\label{fig:hetbet}
\end{figure}

Figures \ref{fig:hetrand}, \ref{fig:hethigh}, and \ref{fig:hetbet} illustrate the effect of heterogeneity on the elasticity of a network.  The interpretation of these Figures is that homogeneous networks have a proclivity for higher levels of elasticity. These include the variations of Watts-Strogatz's small world models, the random models, and the near-regular models. The implications of these results are far reaching where network structure is concerned.  For example, the W-S 2 network is a representative of the random, semi-regular class of topologies where the majority of nodes tend to have a degree close to the average degree.  Therefore, the damage incurred under highest node degree and highest betweenness attacks is comparable.  For example, from Table \ref{tab:attacks} W-S 2 has elasticity scores of $0.164$ and $0.177$ for highest degree and highest betweenness attacks. 

A representative of the deterministic, semi-regular networks, n-r 1 maintains its elasticity under random and high degree attacks. This result can be understood by the almost constant node degree. Thus, random attacks in addition to highest degree attacks, result in similar throughput degradation.  However, as nodes are removed under highest betweenness attacks, core nodes appear and are destroyed.  For n-r 1, elasticity decreases considerably from highest degree attacks to highest betweenness attacks by $35\%$. Thus, although these topologies are sufficiently costly, in addition to the fact that they may fail to capture the properties of some real world networks, their topological structures offer remarkable resilience to attacks.

\begin{figure}[h]
	\centering
		\includegraphics[height=2in]{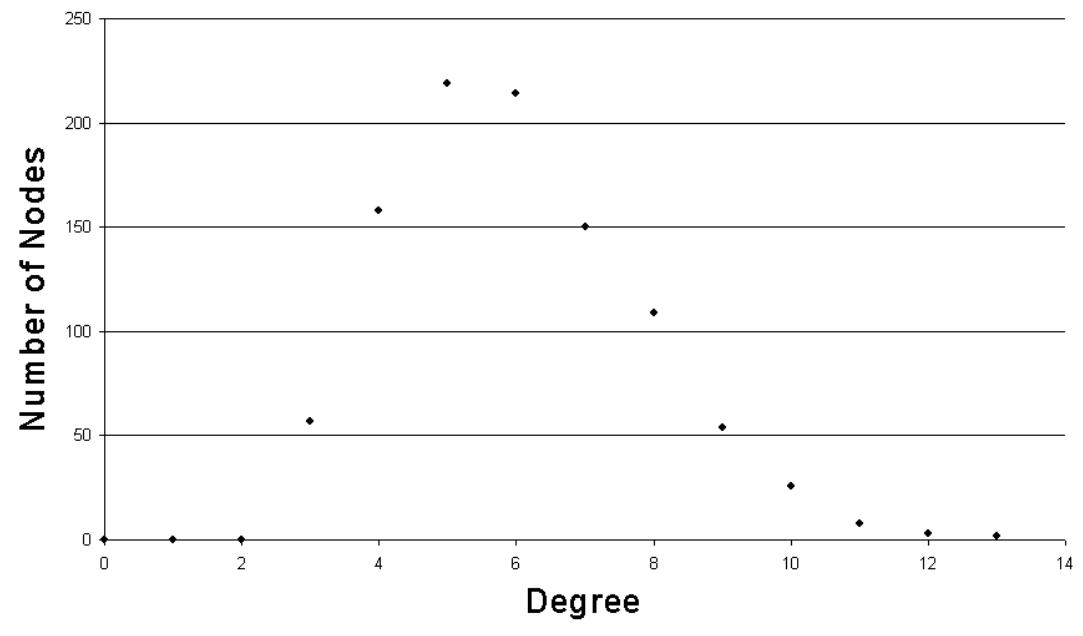}
		 \caption[Degree distribution of Watts-Strogatz 2 network]{Node degree distribution of Watts-Strogatz 2 network}
	\label{fig:nddW-S2}
\end{figure}

\begin{figure}[h]
	\centering
		\includegraphics[height=2in]{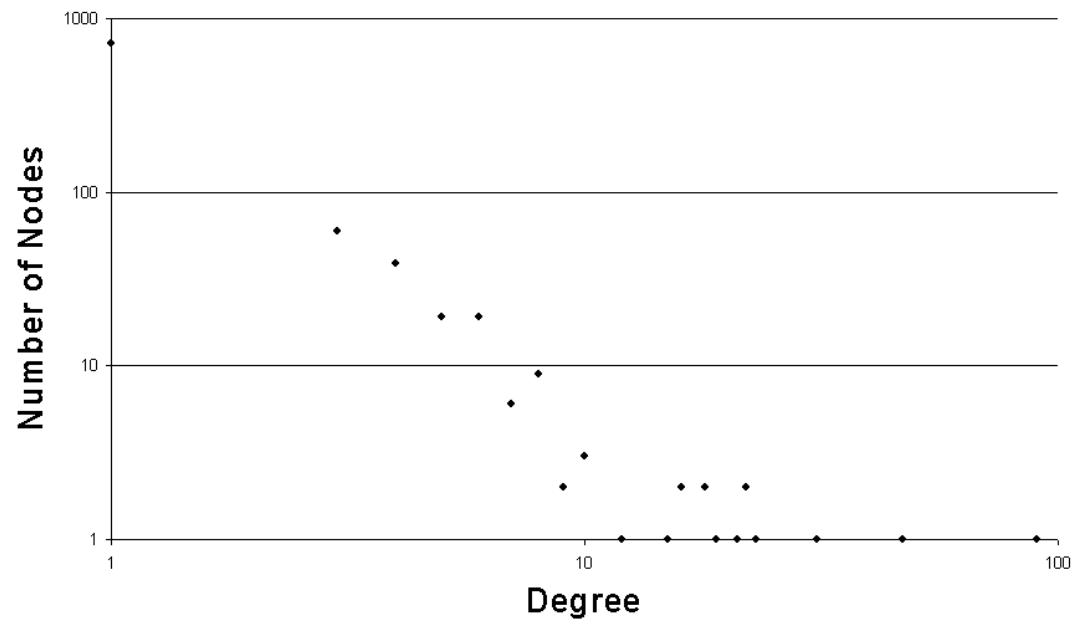}
		 \caption[degree layout of heterogeneous networks]{Node degree distribution of Abilene network}
	\label{fig:nddAbilene}
\end{figure}

Figures \ref{fig:nddW-S2} and \ref{fig:nddAbilene} compare the degree distribution for W-S 2, a representative of the semi-regular class of networks, and Abilene, a representative of the heterogeneous class of networks. As discussed previously, the almost constant degree for the semi-regular class results in high elasticity scores. However, heterogeneous networks span a wide range of degrees and behave differently under attacks. More precisely, based on the ``type" of heterogeneous network under investigation, the impact of highest degree attacks can vary.  On the one hand, networks like Abilene avoid cataclysmic damage under high degree attack because the hubs are located on the periphery of the network and thus, highest degree attack has minimal effect on the overall operation of this network.  However, heterogeneous networks like PA-sparse are severely damaged because the hubs are critical and hold the network together. Thus, homogeneity has far reaching implications in the robustness of networks.

\subsubsection{Correlation Between Elasticity and Characteristic Path Length}
\begin{figure}[h]
	\centering
		\includegraphics[height=2in]{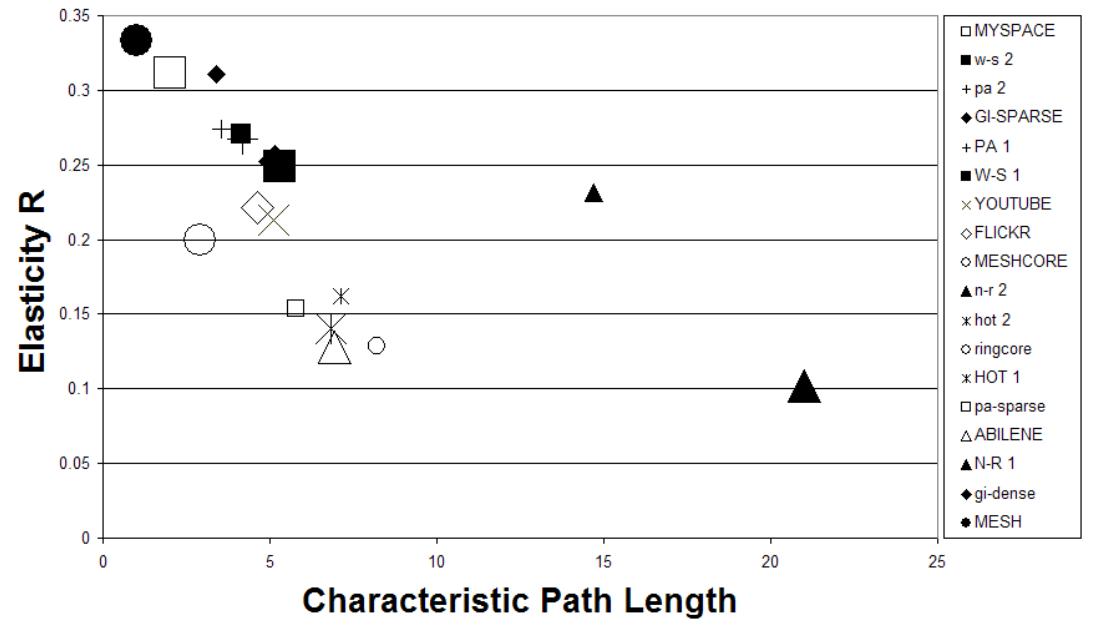}
		 \caption[Convergence of elasticity]{Elasticity R vs characteristic path length for each network in Table \ref{tab:attacks} }
	\label{fig:asprand}
\end{figure}

\begin{figure}[h]
	\centering
		\includegraphics[height=2in]{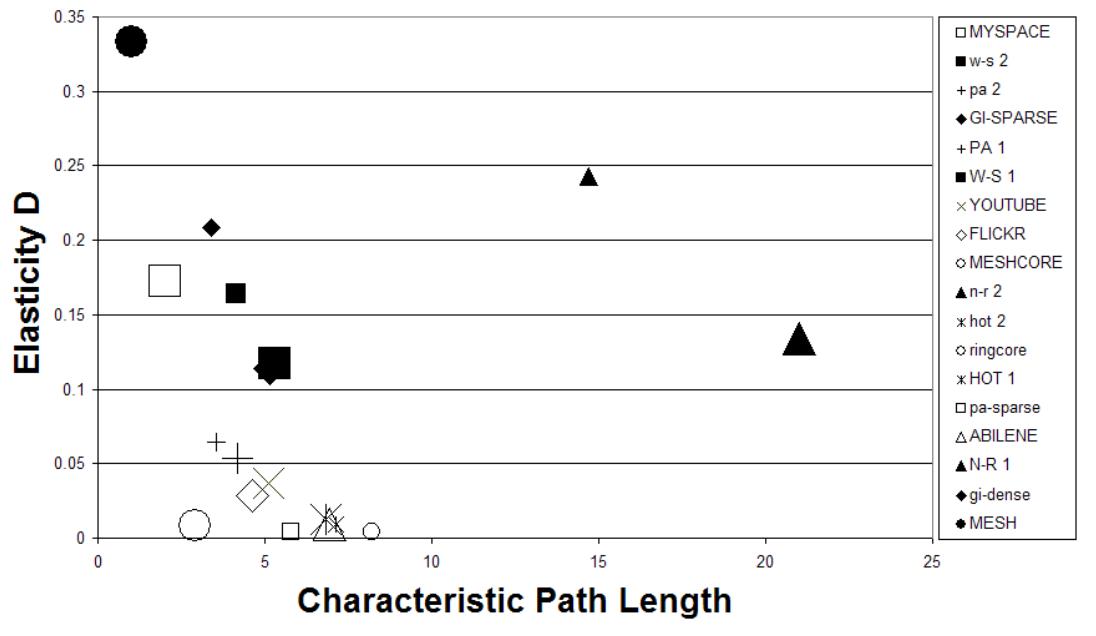}
		 \caption[Convergence of elasticity]{Elasticity D vs characteristic path length  for each network in Table \ref{tab:attacks}}
	\label{fig:asphigh}
\end{figure}

\begin{figure}[h]
	\centering
		\includegraphics[height=2in]{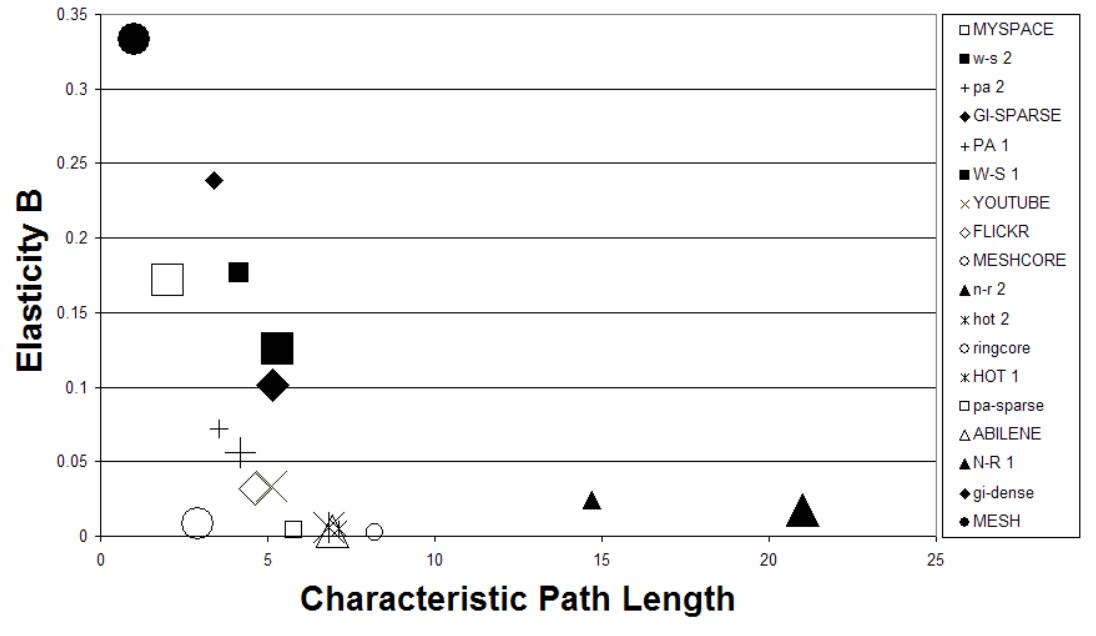}
		 \caption[Convergence of elasticity]{Elasticity B vs characteristic path length for each network in Table \ref{tab:attacks}}
	\label{fig:aspbet}
\end{figure}

The characteristic path length tells the expected distance, in number of hops, from a given source node $s$ to a destination node $t$. Figures \ref{fig:asprand},  \ref{fig:asphigh}, and \ref{fig:aspbet} show that the characteristic path length tends to be negatively correlated with elasticity. This is not a necessary condition as these Figures provide instances where a network with high characteristic path length can have a higher elasticity than a network with a smaller characteristic path length. However, if the number of nodes in a given network is kept constant as the number of links increase, path diversity will eventually increase. As a result, network congestion decreases which ultimately increases elasticity.

\subsection{Elasticity of Networks with Tradeoff Function Applied}

To compensate for the tradeoff between elasticity and number of links, we introduce a tradeoff function $Re\left(G\right)$ that provides robustness with respect to elasticity. For a given network $G$, our robustness measure can be computed as 

\begin{equation}
\centering
\label{eqn:cost} 
 R{e}\left(G\right)=\alpha A + \beta B +  \delta C - \gamma density'  
\end{equation} 

\begin{flushleft}
where $A$, $B$, $C$, represent Elasticity R, Elasticity D, and Elasticity B. $0 \leq \left(\alpha, \beta,\delta, \gamma\right) \leq 1$, $0 \leq density' \leq 1$ and $density' = 1 - e^{-\frac{1}{2}
\frac{\left(M-\left(N-1\right)\right)}{N}}$. The $\frac{1}{2}$ factor determines the rate at which $density'$ changes. $\alpha$, $\beta$,$\delta$, and $\gamma$ are tolerance parameters and as such, represent the tolerance of a network towards random, targeted, and highest betweenness attacks. $M$ is the total number of links and $M-\left(N-1\right)$ represents the number of excess links in a network: these are links which exceed the threshold necessary to obtain $1$ connected component with N nodes. \end{flushleft}

This function facilitates independence for constructing networks based on a projected need.  Thus, a network engineer who envisions persistent, random attacks would consider a high value of $\alpha$. Similarly,  $\beta$ or $\delta$ would dominate where targeted attacks or highest betweenness attacks respectively are predominant.  Moreover, $\gamma$ could be varied based on financial constraints. 

Table \ref{tab:costfunction} depicts the rankings of each topology with their respective number of links and $Re$ scores. For each network, we obtained $Re$ for tolerance values of $\alpha = \beta = \gamma = \delta = 1$. The common tolerance values facilitate an unbiased analysis of robustness by providing equal likelihood of occurrence to each attack strategy. In addition, these rankings represent the case where networks are completely penalized for having excess links and as a result, the structure of the network plays a more significant role to determine the robustness of the network.

\begin{table}
\centering
 \caption{Ranking of networks after implementing the cost function $Re$}
    \begin{tabular}[c]{|c|c|c|
    }        \hline
        {\bf Networks}    &  { \bf Number of links }   & {\bf $Re$}  \\
        \hline
					HOT 2	&	1049	&	0.1519	\\ \hline
					PA-sparse	&	1049	&	0.1374	\\ \hline
					ringcore	&	1000	&	0.1351	\\ \hline
					Abilene	&	896	&	0.1342	\\ \hline
					HOT 1	&	988	&	0.1330	\\ \hline
					W-S 1	&	2000	&	0.0982	\\ \hline
					meshcore	&	1275	&	0.0874	\\ \hline
					YouTube	&	1576	&	0.0828	\\ \hline
					Gi-sparse	&	2009	&	0.0708	\\ \hline
					Flickr	&	1515	&	0.0340	\\ \hline
					PA 1	&	1981	&	-0.0110	\\ \hline
					W-S 2	&	3000	&	-0.0210	\\ \hline
					Gi-dense	&	4505	&	-0.0684	\\ \hline
					near-regular 1	&	1921	&	-0.1206	\\ \hline
					PA 2	&	2964	&	-0.2150	\\ \hline
					near-regular 2	&	3781	&	-0.2561	\\ \hline
					MySpace	&	10976	&	-0.3388	\\ \hline
 
					  \end{tabular}
   
    \label{tab:costfunction}  
   
\end{table}

From this analysis, HOT 2 was the highest ranked network. This network has only $50$ excess links and thus, it virtually avoids the penalty for the existence of excess links.  Furthermore, though it exhibits power-law properties, the hubs are located on the periphery of the network and hence, HOT 2 has an admirable structure against targeted attacks but becomes vulnerable under highest betweenness attacks.  However, considering the values for the tolerance parameters and number of links discussed previously, the HOT 2 network is the most suitable.

\subsection{Tradeoff Between Characteristic Path Length and Heterogeneity}
\begin{figure}[h]
	\centering
		\includegraphics[height=2.5in]{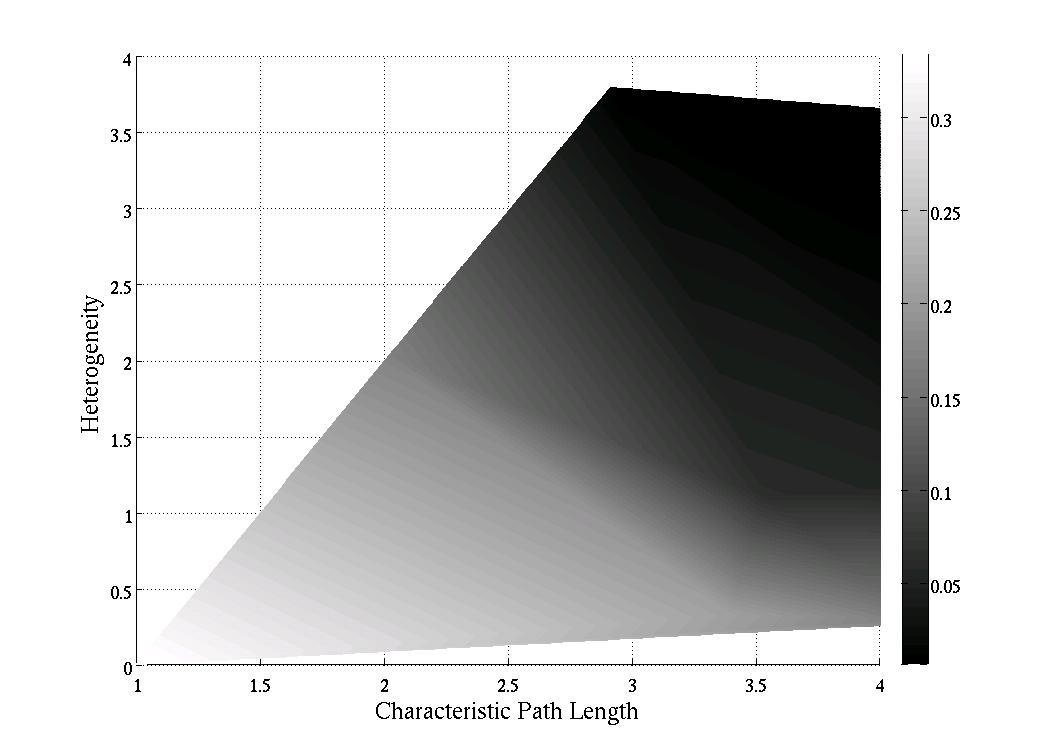}
		 \caption[Elasticity increases as characteristic path length and network heterogeneity decrease]{Elasticity increases as characteristic path length and network heterogeneity decrease }
	\label{fig:hetandcpld}
\end{figure}

The ideal network to provide high elasticity tends to exhibit a low score for heterogeneity and a short characteristic path length. In all networks, the mesh has the shortest characteristic path and the lowest score for heterogeneity and hence, it features the highest elasticity.  However, this high elasticity comes at a very high cost which network designers are unwilling to consider.  For this reason, it is imperative to consider a tradeoff between a short characteristic path length and a low score for heterogeneity. Figure \ref{fig:hetandcpld} shows a plot of heterogeneity against characteristic path length.  The colorbar (to the left) provides the third dimension to this plot of elasticity.  This plot can be interpreted as a decrease in the characteristic path length such that the network becomes more homogeneous increases elasticity. 

\section{Conclusions and Outlook}
\label{conclusions}

This paper endeavors to extract the characteristics of robust complex networks. As our measure for robustness, we used elasticity, which measures the ability of a network to maintain its total throughput under increasing removal of nodes with respective links, and theoretically derived its upper bound of $\frac{1}{3}$. We then illustrated its utility on 18 networks from six different network models under random, highest degree, and highest betweenness attacks and then implemented a tradeoff function which computes robustness with respect to elasticity.

Elasticity is defined and computed under simple assumptions.  As an example, it is dependent on the routing algorithm used, which can perhaps alter current network rankings.   However, elasticity provides benefits which are far-reaching. More precisely, it identifies key characteristics of robust complex networks: A short characteristic path length, low heterogeneity, and strategically located links to facilitate a ``homogeneous" core such that if hubs should be added, they should be placed on the periphery of the network to provide added resilience against targeted attacks. 

For our future work, we intend to incorporate expander graphs in our evaluation and formulate a working definition of the core and periphery to include details about the size and characteristics.  Armed with this knowledge, we seek to combine particular graphs to determine the essential components to increase elasticity.  Finally, we will develop heuristics to build graphs such that elasticity is maximized.  

\section{Acknowlegements}
This research was supported by National Science Foundation (NSF) under award number 0841112. We would like to thank Dr. Robert Kooij for contributing to the success of this work.    

\bibdata
\thebibliography


\end{document}